\documentclass[namedreferences,hyperref,optionalrh]{spr-sola}
\usepackage{graphicx}
\usepackage{adjustbox}
\usepackage{color}
\usepackage{newunicodechar}
\newunicodechar{−}{\textminus}

\newcommand{\aap}{{\it Astron. Astrophys.}}

\newcommand{\apj}{{\it Astrophys. J.}}
\newcommand{\apjl}{{\it Astrophys. J. Lett.}}

\newcommand{\jgr}{{\it J. Geophys. Res.}}

\newcommand{\mnras}{{\it Mon. Not. R. Astron. Soc.}}

\newcommand{\solphys}{{\it Sol. Phys.}}

\chardef\us=`\_

\begin{document}

\begin{frontmatter}
\title{Deep Learning-Enabled Prediction of Geoeffective CMEs Using SOHO and SDO Observations}

\author[addressref={aff1,aff2},email={zy7@njit.edu}]{\inits{Z.}\fnm{Zhaoxin}~\snm{Yan}}
\author[addressref={aff1,aff2},corref,email={wangj@njit.edu}]{\inits{J.T.L.}\fnm{Jason~T.~L.}~\snm{Wang}}
\author[addressref={aff1,aff3,aff4},corref,email={haimin.wang@njit.edu}]{\inits{H.}\fnm{Haimin}~\snm{Wang}}
\author[addressref={aff1,aff3},email={harim.lee@njit.edu}]{\inits{H.}\fnm{Harim}~\snm{Lee}}
\author[addressref={aff1,aff3,aff4},email={ju.jing@njit.edu}]{\inits{J.}\fnm{Ju}~\snm{Jing}}
\author[addressref={aff1,aff3,aff4},email={yan.xu@njit.edu}]{\inits{Y.}\fnm{Yan}~\snm{Xu}}
\author[addressref={aff1,aff2},email={cx4@njit.edu}]{\inits{C.}\fnm{Chunhui}~\snm{Xu}}
\author[addressref=aff4,email={vasyl.yurchyshyn@njit.edu}]{\inits{V.}\fnm{Vasyl}~\snm{Yurchyshyn}}

\address[id=aff1]{Institute for Space Weather Sciences, New Jersey Institute of Technology, University Heights, Newark, NJ 07102, USA}
\address[id=aff2]{Department of Computer Science, New Jersey Institute of Technology, University Heights, Newark, NJ 07102, USA}
\address[id=aff3]{Center for Solar-Terrestrial Research, New Jersey Institute of Technology, University Heights, Newark, NJ 07102, USA}
\address[id=aff4]{Big Bear Solar Observatory, New Jersey Institute of Technology, 40386 North Shore Lane, Big Bear City, CA 92314, USA}

\runningauthor{Yan et al.}
\runningtitle{Deep Learning-Enabled Prediction of Geoeffective CMEs}

\begin{abstract}
Understanding and forecasting the geoeffectiveness 
of a coronal mass ejection (CME)
is crucial for protecting infrastructure in the near-Earth space environment and on Earth. 
In this study, we present a novel fusion model to forecast the geoeffectiveness of CME events.
Our model
combines convolutional neural networks for feature learning and
a prediction network for feature fusion and event classification.
The model is trained by observations from instruments including
the Large Angle Spectroscopic Coronagraph
(LASCO)
on board the Solar and Heliospheric Observatory (SOHO)
and the Atmospheric Imaging Assembly (AIA) 
and Helioseismic and Magnetic Imager (HMI)
on board the Solar Dynamics Observatory (SDO). 
The trained model is then used to predict whether an Earth-reaching CME will
cause a geomagnetic storm and/or
the probability that the CME 
will cause such a storm.
Experimental results 
based on a five-fold cross validation scheme
demonstrate the good performance of
our fusion model, achieving a mean true skill statistic (TSS) score of 0.703
when the model is used as a deterministic prediction tool,
and a mean Brier score of 0.095
when the model is used as
a probabilistic forecasting tool,
where a TSS score of 1 or a Brier score of 0 indicates perfect performance.
This work contributes to forecasting
the causal relationship between Earth-directed CMEs and
geomagnetic storms in 
solar-terrestrial interactions.
\end{abstract}
\end{frontmatter}

\section{Introduction}
\label{S-Introduction} 

Geomagnetic storms, often triggered by
Earth-directed
coronal mass ejections 
\citep[CMEs;][]{2006ApJ...646.1335L,2009EP&S...61..595G,
Vourlidas2019,2025ApJ...981...37Z}, 
pose significant threats to modern technological infrastructure
in the near-Earth space environment and on Earth. 
These disturbances can compromise satellite operations, disrupt global positioning systems, 
interfere with radio communications, and induce harmful currents in power transmission networks. 
The severity of such storms is quantified through various geomagnetic indices, 
with the Dst (Disturbance Storm Time) index serving as a primary indicator, 
where Dst values below $-50$ nT signify geomagnetic storm conditions
\citep{Gonzalez1994, 2022FrASS...9.5880T}. 
Additional indices, including Kp, Ap,
and high-resolution variants such as 
SYM-H and ASY-H, provide complementary storm characterization capabilities \citep{doi:https://doi.org/10.1002/9781118663837.ch2,2006JGRA..111.2202W}.

Recently, 
the rapid advancement of artificial intelligence and machine learning has opened new avenues for forecasting the terrestrial effects of CMEs.
For example, \citet{2019JASTP.19305036B}
adopted a logistic regression model
to predict the geoeffectiveness of a given CME using CME parameters. 
\citet{2022ApJ...934..176P} extended the work of
\citet{2019JASTP.19305036B}
by considering a suite of machine learning methods,
including logistic regression, K-nearest neighbors (KNN), support vector machines (SVMs), feed-forward artificial neural networks (ANNs) and ensemble models, with solar onset parameters.
\citet{2025ApJ...978...66Y} 
developed a feature dimension reduction method
in conjunction with a KNN model to predict the geoeffectiveness of CMEs using multiple CME features.
\citet{2024ApJ...972...52Y}
investigated the geoeffectiveness of
a CME and its dependence on solar wind conditions
using an SVM model.
\citet{2023ApJ...952..111T}
predicted geomagnetic events,
also using solar wind data with multiple ANN models.
\citet{Guastavino_2024}
formulated the prediction of geomagnetic disturbances
as a binary classification problem and
employed
a long short-term memory recurrent neural network, 
together with in situ measurements of solar wind plasma and magnetic field,
to solve the problem.
\citet{Hu_2023} designed multi-fidelity boosted neural networks, trained by solar wind parameters,
to predict the Dst index.
\citet{Liu2024} adopted a temporal convolutional network with integrated gradients to predict Dst.
\citet{2021RemS...13.1738F} used solar images to
predict the geoeffectiveness of CMEs
with an attention-assisted convolutional neural network (CNN).
Because only solar observations were used without handpicked features, their
tool has the potential for making near-real-time forecasts.

In this study, we present a novel
fusion model that
combines convolutional neural networks for feature learning and
a prediction network for feature fusion and event classification,
to predict geoeffective CMEs.
Specifically, we use observations collected by
the Atmospheric Imaging Assembly 
\citep[AIA;][]{2012SoPh..275...17L} 
and Helioseismic and Magnetic Imager 
\citep[HMI;][]{2012SoPh..275..207S}
on board the Solar Dynamics Observatory 
\citep[SDO;][]{2012SoPh..275....3P}
as well as
the Large Angle Spectroscopic Coronagraph
\citep[LASCO;][]{1995SoPh..162..357B}
on board the Solar and Heliospheric Observatory 
\citep[SOHO;][]{1995SoPh..162....1D}
to train and test our fusion model.
The model predicts whether 
an Earth-reaching CME will
cause a geomagnetic storm
and/or
the probability that the CME 
will cause such a storm,
where the 
storm is defined as a disturbance
of the Earth's magnetosphere during which the minimum Dst value
is lower than $-50$ nT 
\citep{Gonzalez1994, 2022FrASS...9.5880T}.
The fusion model achieves a mean true skill statistic (TSS) score of 0.703
when used as a deterministic prediction tool,
and a mean Brier score of 0.095 when used as
a probabilistic forecasting tool
based on a five-fold cross validation scheme,
where a TSS score of 1 or a Brier score of 0 indicates perfect predictions.

The remainder of this paper is organized as follows. Section \ref{sec:data} describes the observational data and pre-processing procedures used in our study. 
Section \ref{sec:methods} presents the architecture and implementation details of our fusion model. 
Section \ref{sec:results} reports the experimental results obtained by evaluating the fusion model
from the perspectives of deterministic prediction and 
probabilistic forecasting.
Section \ref{sec:conc} 
presents a discussion and 
concludes the paper.

\section{Data}
\label{sec:data}

\begin{figure}
   \centering
   \includegraphics[width=0.8\textwidth]{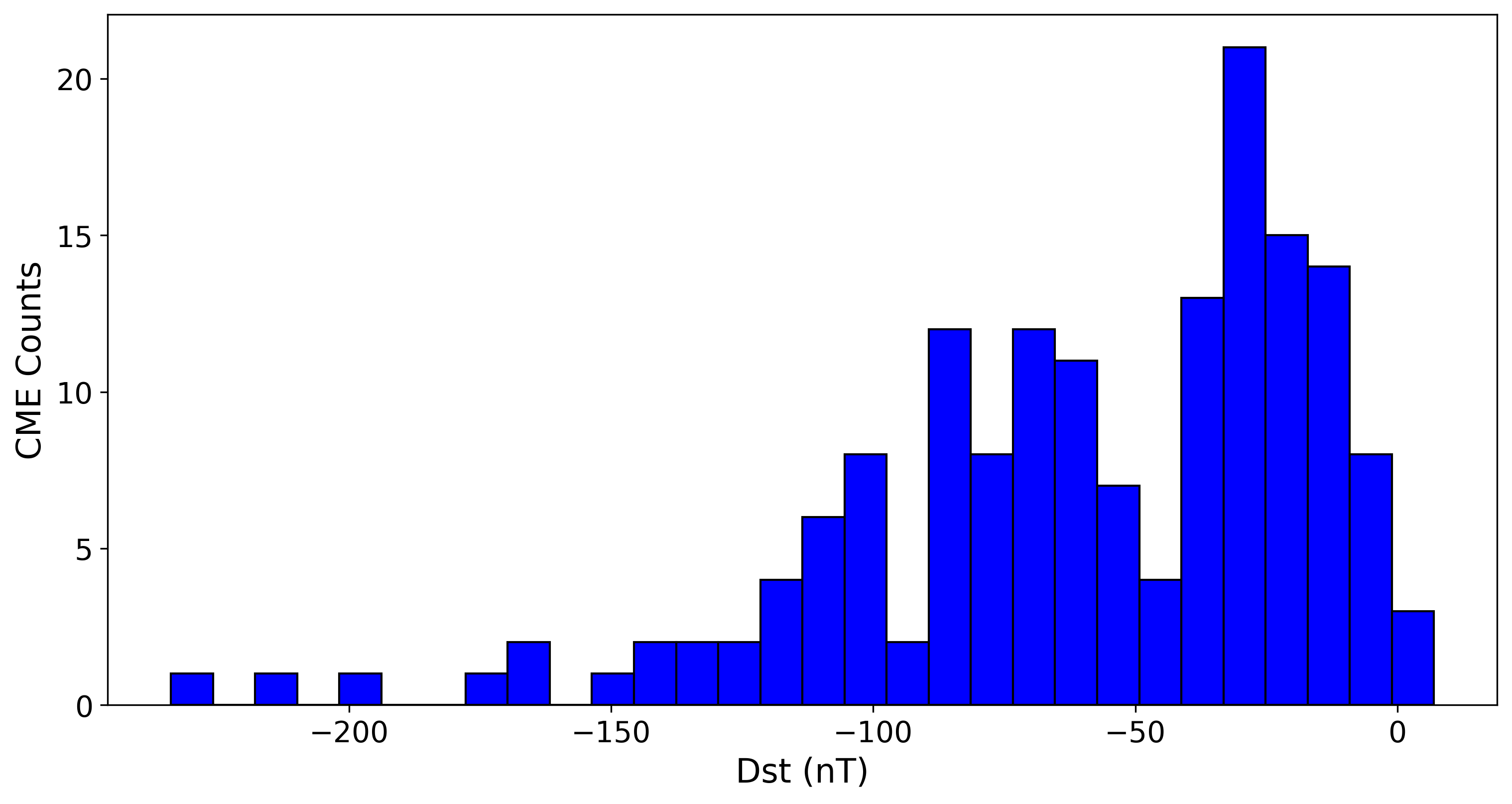}
   \caption{Distribution of the Dst index values for the 
   164 Earth-reaching CME events in our dataset.} 
   \label{fig:dst_dist}
\end{figure}

\begin{figure}
   \centering
   \includegraphics[width=0.8\textwidth]{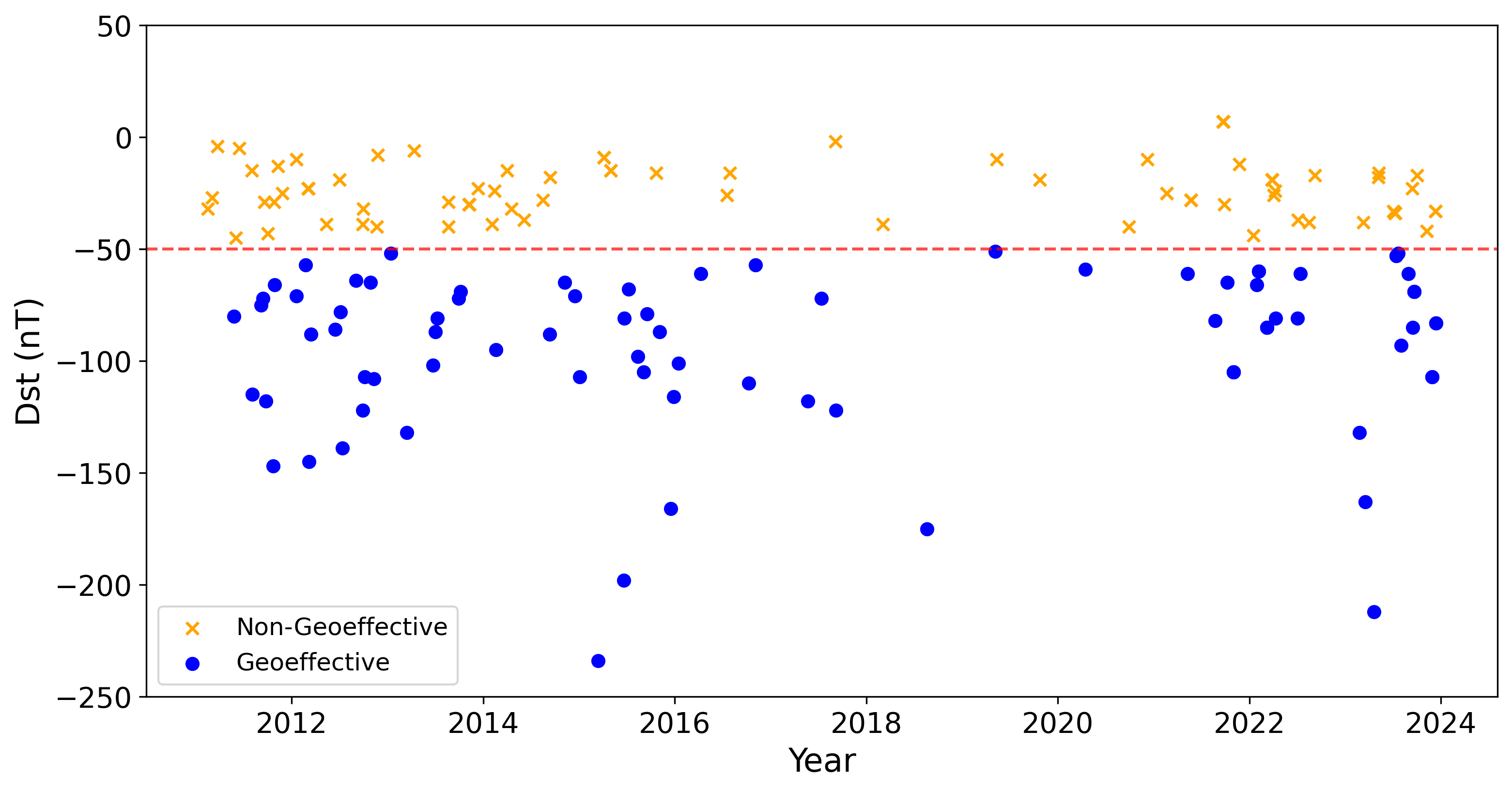}
   \caption{Breakdown of the 
   geoeffective (blue)
   and non-geoeffective (orange)
   CME events in our dataset.}
   \label{fig:dst_time}
\end{figure}

The main data source used in our study is the
list of interplanetary CMEs maintained by \citet{2010SoPh..264..189R},
referred to as the RC list.
We selected 164 Earth-reaching CMEs that have LASCO observing time
recorded in the RC list. 
These CME events spanned the period from
2011 to 2024.
All selected CMEs were associated with Dst values in the RC list.
Figure \ref{fig:dst_dist} 
shows the distribution of the Dst values
for the 164 CMEs.
A CME is geoeffective if its associated Dst value is lower than $-50$ nT \citep{Gonzalez1994, 2022FrASS...9.5880T}.
As a result, there were
86 geoeffective CMEs and 78 non-geoeffective CMEs.  
Figure \ref{fig:dst_time} shows the breakdown of these CME events.
The lower the Dst values, the stronger the corresponding geomagnetic storms.

\begin{figure}
    \centering
    \includegraphics[width=0.8\textwidth]{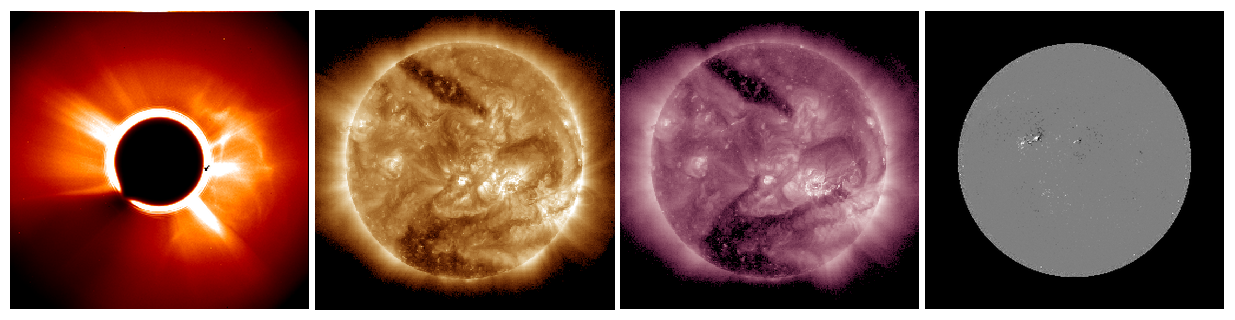}
    \caption{Solar observations on the CME event that occurred on
14 March 2015 UT.
Shown from left to right are a LASCO C2 image, 
an AIA 193~\AA{} image,
an  AIA 211~\AA{} image, 
and a full-disk HMI line-of-sight magnetogram.}
    \label{fig:soho_obs}
\end{figure}

When training and validating our fusion model, we used
solar observations from multiple instruments,
including SOHO/LASCO, SDO/AIA with two channels
193~\AA{} and 211~\AA{},
and SDO/HMI.
Figure \ref{fig:soho_obs} 
shows the solar observations on the CME event that occurred on
14 March 2015 UT, 
which are, from left to right, 
LASCO C2, 
AIA 193~\AA{},
AIA 211~\AA{},
and 
HMI.
As in the literature \citep{2019ApJ...881...15W,
2023ApJ...958L..34A},
we used base-difference
images for LASCO C2 to improve model performance.
LASCO C2 images capture the early outward propagation of CMEs in the corona and provide onset signatures and initial speed estimates 
\citep{1995SoPh..162..357B}. 
AIA images show coronal dimmings and flare-related brightenings in the low corona that are strongly associated with CME initiation
\citep{2018ApJ...863..169D,2019ApJ...874..123D},
and they also reveal coronal holes, which are the sources of high-speed solar wind streams that can interact with CMEs and affect their geoeffectiveness 
\citep{2009JGRA..114.0A22G}.
HMI magnetograms provide the photospheric magnetic field context, including shear, polarity inversion lines, and sigmoidal structures, which are known to increase the probability of eruptions
\citep{2006ApJ...644.1258F,2019LRSP...16....3T}.
 Combining these observations gives a multimodal view of both the magnetic environment and the coronal response, improving the ability to assess CME potential and forecast their geoeffectiveness.

Each CME event in our dataset had a LASCO observing time stamp, denoted $t$, recorded in the RC list.
Following \citet{2024SoPh..299..159A},
we collected LASCO C2 images
10 minutes before $t$ and 4 hours after $t$, averaging
21 images per CME event, with a total of 3,438 
LASCO C2 images.
In addition, we collected
AIA 193~\AA{} images
and AIA 211~\AA{} images, with a cadence of 1 hour,
in the time window from 4 hours before $t$ to $t$,
totaling 1,295 AIA images.
Finally, we collected
HMI magnetograms, also with a cadence of 1 hour,
in the time window from 3 hours before $t$ to $t$,
totaling 471 HMI magnetogram images.
These temporal ranges were chosen
 on the basis of the literature \citep{2024SoPh..299..159A}
and also our own experimental study.
We have tested larger temporal ranges, which incurred longer 
execution times
without improving performance,
and smaller temporal ranges, which yielded worse performance.
Missing images in the time windows were excluded from the study.
The dataset contained 5,204 images in total.

We then used a stratified 80:20 split 
on our CME event dataset
for model training and testing.
Specifically, we randomly selected 80\% 
of the 86 geoeffective CME events in the dataset,
which yielded 69 events,
and 80\% of the 78 non-geoeffective CME events in the dataset,
which yielded 62 events, for model training. 
The remaining 20\% from each class, 
including 17 geoeffective and 16 non-geoeffective events, 
were used for model testing. 
Thus, we maintained the
same class proportions 
(geoeffective vs. non-geoeffective)
in both training and test sets
as the original event dataset.
All images associated with an event were exclusively assigned 
to the same set, either training or test,
to prevent data leakage.
The training set and the test set are disjoint and, therefore,
our trained model can make predictions on test data that it has never seen during training.
To monitor model performance and prevent overfitting during training, 
we further selected 12\% of the training 
events
from each class and used the selected events
for model validation.

\section{Methodology} 
\label{sec:methods}

\begin{figure}
    \centering
    \includegraphics[width=\textwidth]{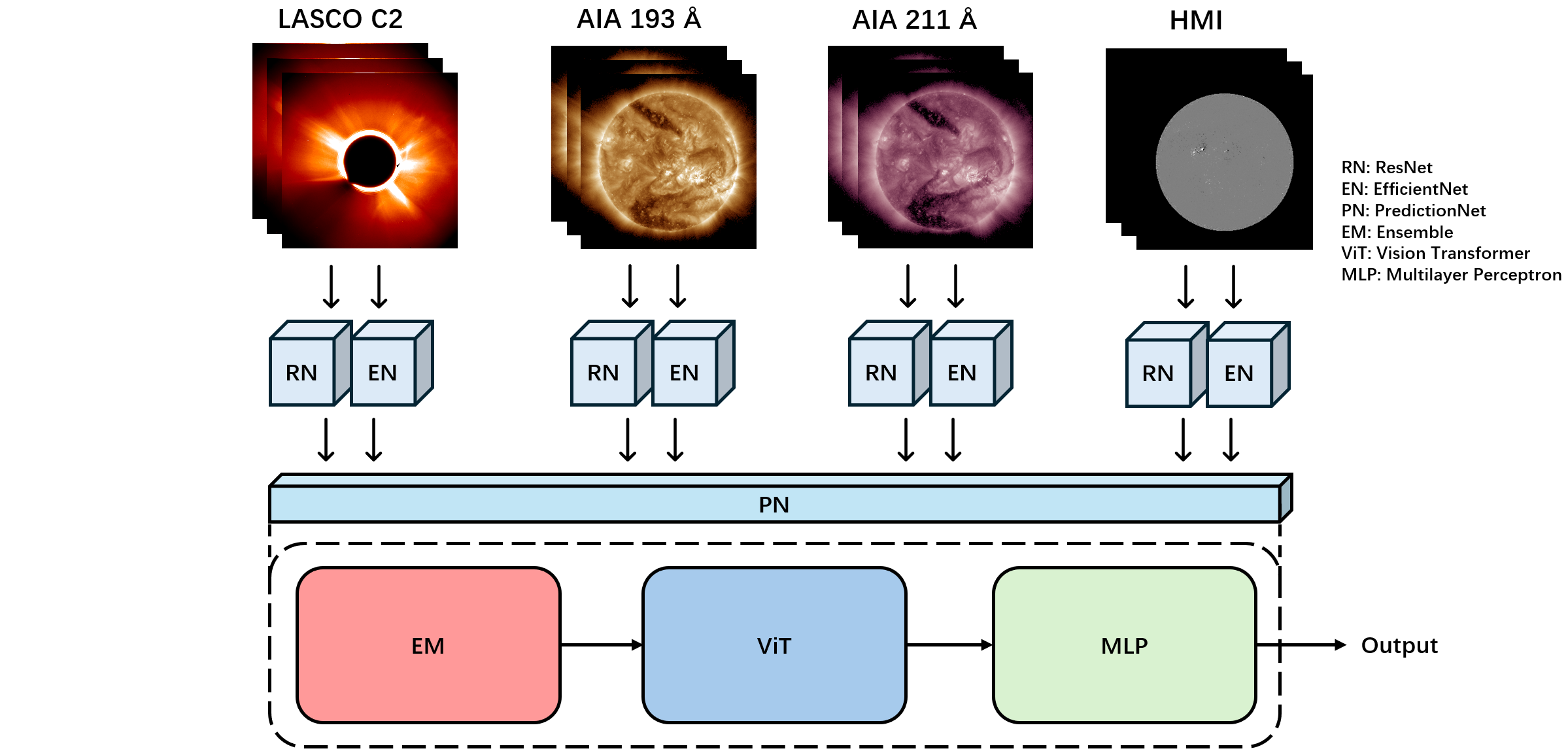}
    \caption{Overall architecture of our fusion model.
The model accepts, as input, a CME event
represented by images of four types
(LASCO C2, AIA 193~\AA{}, AIA 211~\AA{} and HMI),
and assigns a pair of networks, ResNet (RN) and EfficientNet (EN),
to learn and extract features from the input images of each type.
These learned features are then sent to 
our PredictionNet (PN), which is composed of
three modules: 
ensemble (EM), 
vision transformer (ViT) and
multilayer perceptron (MLP),
for feature fusion and event classification.
PN outputs ``1'' indicating that the input CME event is positive/geoeffective
or
``0'' indicating that the input CME event 
is negative/non-geoeffective
when our model is used as a deterministic prediction tool.
When our model is used as a probabilistic forecasting tool,
the output is a probability for the input CME event to be
geoeffective.}
    \label{fig:fusion_architecture}
\end{figure}

Figure \ref{fig:fusion_architecture}
shows the architecture of our fusion model.
The input of
the fusion model consists of solar images 
associated with a CME event.
Letting the LASCO observing time of the CME event be $t$,
these solar images, which represent the CME event, include:
21 LASCO C2 images 
in the period from $t$ - 10 min to $t$ + 4 hr,
4 AIA 193 \AA{} images and 
4 AIA 211 \AA{} images 
in the period from $t$ - 4 hr to $t$,
and 3 HMI magnetograms 
in the period from $t$ - 3 hr to $t$.

We employ two convolutional neural networks (CNNs), specifically
ResNet50 \citep{He2016} and
EfficientNet-B0 \citep{Tan2019},
for feature learning and a vision transformer
\citep{DBLP:conf/iclr/DosovitskiyB0WZ21}
for image classification based on learned features.
CNNs are well known for their ability to learn useful features from raw pixel data.
ResNet50 and EfficientNet-B0 are widely used CNNs due to their balance of performance and computational cost.
Both ResNet50 and EfficientNet-B0 have been trained in ImageNet
\citep{Deng2009} and are adapted to our task through
transfer learning \citep{DBLP:journals/tkde/PanY10}.
Specifically, we load the pre-trained weights of ResNet50
and EfficientNet-B0 to our
ResNet and EfficientNet models, respectively, and fine-tune
the models using the training and validation images at hand.
We assign a pair of ResNet and EfficientNet to each data type
(LASCO C2, AIA 193~\AA{}, AIA 211~\AA{} and HMI),
where the two CNNs are
used to learn features from images of
the respective data type.

For each data type, 
ResNet processes each image of that data type, individually and separately,
through its deep residual architecture
to produce a 2,048-channel feature map
with a spatial dimension of $7 \times 7$ pixels.
The feature maps are intermediate representations produced by the convolutional layers in ResNet. 
Specifically, each data type has multiple images and hence is associated with
multiple feature maps. 
For example, for
the 21 LASCO C2 images, we obtain 21 feature maps.
These 21 feature maps are averaged
to obtain a single representative 2,048-channel feature map with a spatial dimension of
$7 \times 7$ pixels.
The average feature map preserves important temporal dynamics and 
fine-grained spatial features
in the 21 LASCO C2 images.
Similarly,
for the 4 AIA 193 \AA{} images
(4 AIA 211 \AA{} images, 
3 HMI magnetogram images, respectively),
we obtain 4 
(4, 3, respectively) feature maps, which are
averaged to obtain a single representative 2,048-channel feature map with a spatial dimension of
$7 \times 7$ pixels for the
4 AIA 193 \AA{} images
(4 AIA 211 \AA{} images, 
3 HMI magnetogram images, respectively).
Simultaneously, EfficientNet uses its compound-scaled convolutional layers
to produce a representative, average 1,280-channel feature map
with the same spatial resolution of $7 \times 7$ pixels
for the multiple images of each data type.
The representative, average feature maps of each data
type are then sent to a prediction network, or PredictionNet (PN), which
is composed of three modules:
ensemble (EM), vision transformer (ViT) and multilayer perceptron (MLP).

For each data type, the EM module combines the two
representative average feature maps produced by ResNet and EfficientNet,
respectively,
through channel concatenation and attention mechanisms
to create a unified
feature map of $512 \times 7 \times 7$ pixels
that preserves spatial information.
This process results in four feature maps, one for
each data type.
Next, the EM module learns weights among the four feature maps to combine them to produce a final 512-dimensional feature map,
which is projected
onto a 768-dimensional space through a linear transformation
to match the internal dimensionality requirements of the
subsequent ViT module.

ViT is well known for its ability to classify images effectively.
ViT transforms the processed 768-dimensional feature map
into an optimized 512-dimensional feature vector.
The MLP module has two fully connected layers: the first layer reduces dimensionality to 256 with ReLU activation, followed by dropout regularization with a dropout rate of 0.3 to prevent overfitting, and a final sigmoid-activated layer,  
which produces a probability for the input CME event to be
geoeffective.
To obtain deterministic predictions, we implement a probability threshold of 0.6 in our fusion model.
CME events with predicted probabilities greater than or equal to the threshold are classified as geoeffective, while CME events with predicted probabilities below the threshold are classified as non-geoeffective.
This threshold value was empirically optimized to balance recall and precision in our validation dataset.
Specifically, the threshold of 0.6 was determined through a grid search on the validation set
containing 9 geoeffective and 8 non-geoeffective CME events. 
We tested threshold values in the range 
between 0.3 and 0.7
with a step size of 0.05 and selected the threshold that maximized the 
true skill statistic (TSS),
which will be defined in Section \ref{sec:setup}. 
TSS provides a balanced assessment of predictive skills that accounts for both the hit rate and the false alarm rate, independent of the class distribution \citep{Bloomfield_2012}. The optimization was performed exclusively on the validation set to avoid data leakage from the test set. 

\begin{figure}
    \centering
    \includegraphics[width=\textwidth]{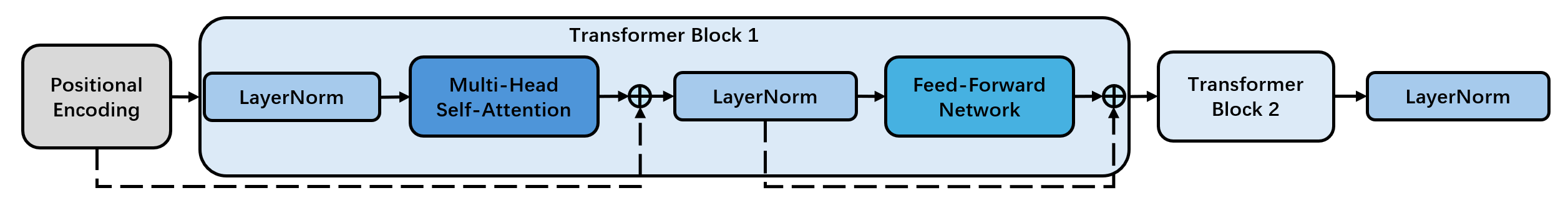}
    \caption{Architecture details of the vision
    transformer (ViT), which consists of multi-head self-attention layers, 
    feed-forward networks, and residual connections for enhanced feature representation.}
    \label{fig:ViT_block}
\end{figure}

Figure \ref{fig:ViT_block} details the ViT module,
where we load the pre-trained weights
of \citet{DBLP:conf/iclr/DosovitskiyB0WZ21}
into the module.
The input 768-dimensional feature map 
with a spatial resolution of $14 \times 14$ pixels
is first partitioned into 196 non-overlapping patches
by a positional encoding layer.
This creates a sequence of 196 patch embeddings, 
each represented as a 768-dimensional vector.
These patch embeddings are then processed through 
two sequential transformer encoder blocks, 
each containing a multi-head self-attention layer
with 12 attention heads and 
a position-wise feed-forward network 
utilizing the GELU activation function.
Residual connections, denoted by $\oplus$, and layer normalization ensure stable gradient flow and feature preservation throughout the transformer processing pipeline.
The transformer blocks process the sequence of 196
patches, learning contextual relationships between the patches through 
the self-attention layers and feed-forward networks.
Following transformer enhancement, the input 768-dimensional feature map undergoes a final layer normalization and is projected into an optimized
512-dimensional feature vector.
Table \ref{tab:fusion_config} and
Table \ref{tab:fusion_params}
summarize the configuration and parameter details, respectively, of
our fusion model.

\begin{table}[ht]
\caption{Configuration Details of the Proposed Fusion Model}
\label{tab:fusion_config}
\begin{tabular}{lcccc}
\hline
Component & Configuration & Regularization & Activation & Output \\
\hline
ResNet & 50 Layers & Batch Norm. & ReLU & $2,048 \times 7 \times 7$ \\
EfficientNet & 38 Layers & Batch Norm. & SiLU & $1,280 \times 7 \times 7$ \\
EM Module & Conv2D + Attn. & Batch Norm. & ReLU & $512 \times 7 \times 7$ \\
ViT Pos. Encoding & Linear Proj. & $-$ & $-$ & $196 \times 768$ \\
ViT Trans. Block 1 & 12 Attn. Heads & Layer Norm. & GELU & $196 \times 768$ \\
ViT Trans. Block 2 & 12 Attn. Heads & Layer Norm. & GELU & $196 \times 768$ \\
MLP Dense Layer & Fully Connected & $-$ & ReLU & 256 \\
MLP Output Layer & Fully Connected & Dropout (0.3) & Sigmoid & 2 \\
\hline
\end{tabular}
\end{table}

\begin{table}
\caption{Parameter Analysis for the Proposed Fusion Model}
\label{tab:fusion_params}
\begin{tabular}{lcc}
\hline
Architecture Component & Layer Count & Trainable Parameters \\
\hline
ResNet& 50 & 23.51M \\
EfficientNet & 38 & 4.01M \\
EM Module & 3 & 1.68M \\
ViT Transformer Enhancement & 2 Blocks & 14.37M \\
ViT Other Layers & 3 & 0.89M \\
MLP Layers & 2 & 0.20M \\
Whole Fusion Model & $-$ & 44.66M \\
\hline
\end{tabular}
\end{table}

\section{Experiments and Results} 
\label{sec:results}
\subsection{Experimental Setup} 
\label{sec:setup}

As mentioned in Section \ref{sec:data},
we implemented a stratified data partitioning scheme
by dividing the dataset at hand into training and test sets
using an 80:20 ratio where 
approximately 12\% of the training data from each class were selected for model validation.
There were 164 CMEs in our dataset, 
with 86 being geoeffective and 78 being non-geoeffective.
As a result, the training set had 69 geoeffective CMEs and 62 non-geoeffective CMEs where
 9 geoeffective CMEs and 8 non-geoeffective CMEs
 are used for model validation,
and the test set had 17 geoeffective CMEs and 16 non-geoeffective CMEs.

When evaluating our fusion model
for deterministic prediction,
we adopted five complementary performance metrics:
recall, precision, accuracy, 
F1, and true skill statistic (TSS).
Given a CME event $E$, 
$E$ is a true positive (TP) 
when the model predicts that $E$ is positive
(i.e. geoeffective)
and $E$ is indeed positive.
$E$ is a true negative (TN)
when the model predicts that $E$ is negative 
(i.e. non-geoeffective)
and $E$ is indeed negative.
$E$ is a false positive (FP)
when the model predicts that $E$ is positive
while $E$ is actually negative.
$E$ is a false negative (FN)
when the model predicts that $E$ is negative
while $E$ is actually positive.
When the context is clear, we also use
TP, TN, FP and FN to
represent the total number of
true positives, true negatives, false positives, and false negatives, respectively.
The five performance metrics are mathematically
defined as follows:
\begin{equation}
\mathrm{Recall} = \frac{\mathrm{TP}}{\mathrm{TP} + \mathrm{FN}},
\end{equation}
\begin{equation}
\mathrm{Precision} = \frac{\mathrm{TP}}{\mathrm{TP} + \mathrm{FP}},
\end{equation}
\begin{equation}
\mathrm{Accuracy} = \frac{\mathrm{TP} + \mathrm{TN}}{\mathrm{TP} + \mathrm{FP} + \mathrm{TN} + \mathrm{FN}},
\end{equation}
\begin{equation}
\mathrm{F1} = \frac{2 \times \mathrm{TP}}{2 \times \mathrm{TP} + \mathrm{FP} + \mathrm{FN}},
\end{equation}
\begin{equation}
\mathrm{TSS} = \frac{\mathrm{TP}}{\mathrm{TP} + \mathrm{FN}} - \frac{\mathrm{FP}}{\mathrm{FP} + \mathrm{TN}}.
\end{equation}

F1 combines precision and recall into a single value, providing a balanced measure of a model's performance.
TSS accounts for both the hit rate (recall) and the false alarm rate, providing a balanced assessment of predictive skills independent of the class distribution, as indicated previously.
The F1 score ranges from 0 to 1, while
the TSS score ranges from $-1$ to 1,
with a value of 1 indicating perfect model performance for both metrics.

In addition to the fusion model,
we consider its three component networks alone, used as baseline methods.
The ResNet baseline takes solar images as input 
from all four data types 
(LASCO C2, 
AIA 193 \AA{}, 
AIA 211 \AA{}, 
and HMI),
which are first processed individually 
within their respective time windows.
The resulting feature maps 
for each data type
are averaged 
to obtain a single representative 2,048-channel feature map with a spatial dimension of
$7 \times 7$ pixels.
Each data type yields an average feature map.
The four average feature maps are then concatenated and fed into a multilayer perceptron (MLP) classifier, which has the same architecture as that used in the fusion model, to produce the output.
The EfficientNet baseline works similarly.
For the ViT baseline, the multiple 
images of each data type 
are first averaged to obtain a single representative image for each data type, respectively.
Each average image is resized to 
$112 \times 112$ pixels.
The four resized average images are then concatenated into  
 a $224 \times 224$ image arranged in a 
 $2 \times 2$ grid.
 Next, the concatenated image is partitioned into $14 \times 14 = 196$ non-overlapping 
patches, each having $16 \times 16$ pixels.
Each patch is linearly projected onto a 
768-dimensional embedding vector, 
which matches the input dimensionality of the ViT baseline.
These patch embeddings are then processed by two transformer encoder blocks 
with the same architecture as those shown in Figure \ref{fig:ViT_block},
followed by an MLP classifier, which has the same architecture as that used in the fusion model, to produce the output.
All three baseline methods and the proposed fusion model use the same training, validation, and test sets, as well as identical optimization settings, including the learning rate, batch size, number of epochs, and loss function, to ensure a fair and controlled comparison between the different models.

\subsection{Deterministic Prediction Results}

\begin{figure}
    \centering
    \includegraphics[width=0.6\textwidth]{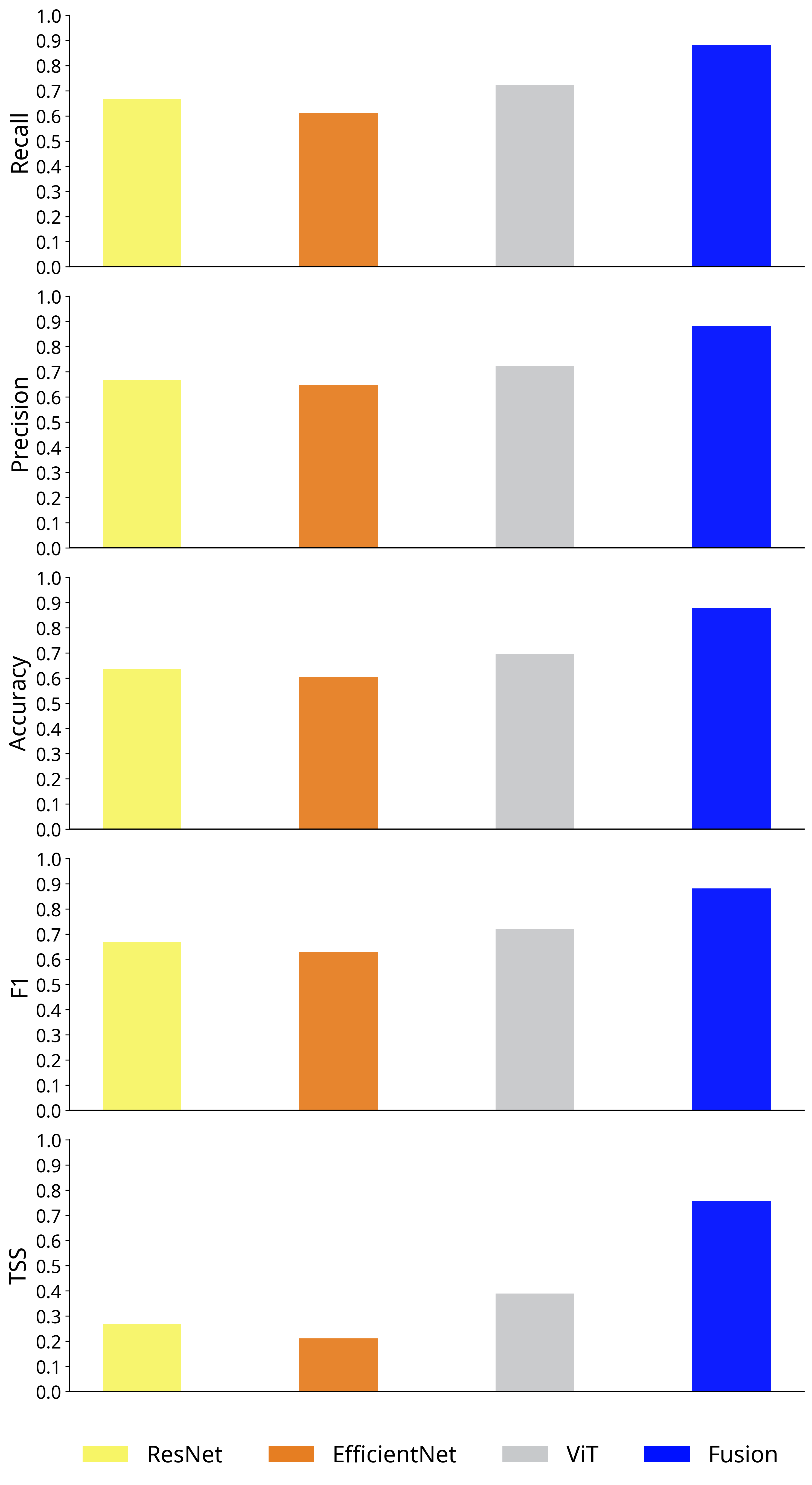}
    \caption{Comparison of our fusion model with its three component networks
    based on the 80:20 train-test split
    described in Section \ref{sec:data}
    when all the four tools are used for deterministic prediction.
    The fusion model achieves the best performance 
    with a TSS score of 0.757.} 
    \label{fig:model_metrics}
\end{figure}

We conducted a series of experiments to evaluate the performance of the
proposed fusion model and its component networks
(i.e., the three baseline methods).
Figure \ref{fig:model_metrics} presents the results.
It can be seen in Figure \ref{fig:model_metrics} that
the ViT baseline outperforms the other two component networks
(ResNet and EfficientNet).
Overall, the fusion model is the best,
achieving a TSS score of 0.757.
This happens probably because
the fusion model combines the superior capabilities of convolutional
neural networks in representation/feature learning and
those of ViT for image recognition/classification.
Previously, 
\citet{2024SoPh..299..159A}
used different data and algorithms to predict
geoeffective CMEs
with a slightly lower TSS of 0.714,
though a direct comparison between 
our work and
Alobaid et al.'s work
is not possible because the 
algorithms and datasets used by
the two works differ.

\begin{figure}
   \centering
   \includegraphics[width=0.4\textwidth]{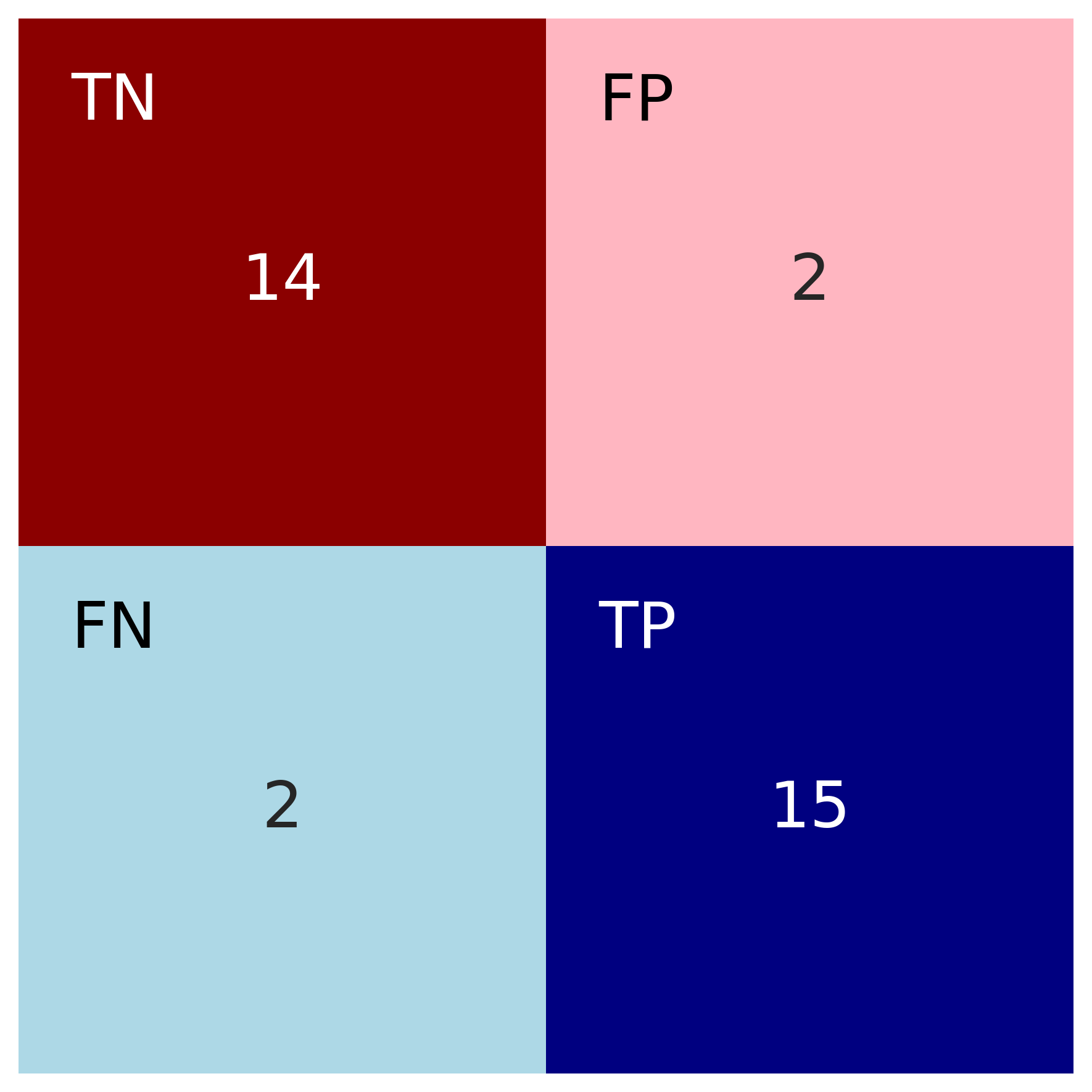}
   \caption{The confusion matrix obtained by our fusion model used for
deterministic prediction
    based on the 80:20 train-test split described in Section \ref{sec:data}.}
   \label{fig:conf_matrix}
\end{figure}

Figure \ref{fig:conf_matrix} presents the confusion matrix
obtained by our fusion model.
The confusion matrix provides a breakdown
of errors that occur when the fusion model makes predictions on the test set. 
There are 33 CME events in the test set,
with 17 being geoeffective/positive and 16 being 
non-geoeffective/negative.
Among the 17 positive CME events,
the fusion model correctly predicts 15 of them (TP = 15)
and incorrectly predicts 2 of them to be negative, although
the two are actually positive (FN = 2).
Among the 16 negative CME events,
the fusion model correctly predicts 14 of them (TN = 14)
and incorrectly predicts 2 of them to be positive, although
the two are actually negative (FP = 2).

To ensure the statistical reliability of our performance assessment, we also performed a five-fold cross-validation experiment using stratified sampling to maintain consistent class distributions across all folds. 
Specifically, 
our dataset of 164 CME events was partitioned into five subsets of approximately equal size, each containing roughly 17 geoeffective and 16 non-geoeffective CMEs. The five-fold cross-validation was applied to all models, including the 
proposed fusion model and 
its component networks (i.e., the three baseline methods).
With the five-fold cross validation,
our fusion model achieved 
a mean TSS of 
 0.703,
 in reference to
 the mean TSS of 
 0.673 in
 \citet{2024SoPh..299..159A}.
 The ResNet baseline achieved
 a mean TSS of 
 0.248, 
 the EfficientNet baseline achieved 
 a mean TSS of 
  0.195, 
  and the ViT baseline achieved 
  a mean TSS of 0.363.
  Table \ref{tab:cv_deterministic}
  presents the detailed results
  including the mean scores and standard deviations
  of all performance metrics, with
  the best metric values highlighted in boldface.
  The five-fold cross validation results again demonstrate the superiority of the proposed fusion model
  over the baseline methods.

\begin{table}
\caption{Five-Fold Cross Validation Results of the Four Deterministic Prediction Models}
\label{tab:cv_deterministic}
\setlength{\tabcolsep}{5pt}
\begin{tabular}{@{}lccccc@{}}
\hline
Model & Recall & Precision & Accuracy & F1 & TSS \\
\hline
ResNet       & 0.620$\pm$0.137 & 0.625$\pm$0.111 & 0.624$\pm$0.028 & 0.622$\pm$0.124 & 0.248$\pm$0.055 \\
EfficientNet & 0.565$\pm$0.174 & 0.604$\pm$0.149 & 0.598$\pm$0.030 & 0.584$\pm$0.162 & 0.195$\pm$0.060 \\
ViT          & 0.674$\pm$0.089 & 0.684$\pm$0.072 & 0.681$\pm$0.024 & 0.679$\pm$0.081 & 0.363$\pm$0.048 \\
Fusion       & {\bf 0.819}$\pm$0.035 & 
    {\bf 0.876}$\pm$0.030 & 
    {\bf 0.851}$\pm$0.015 & 
    {\bf 0.847}$\pm$0.033 & 
    {\bf 0.703}$\pm$0.030 \\
\hline
\end{tabular}
\end{table}

\begin{figure}
    \centering
    \includegraphics[width=0.85\textwidth]{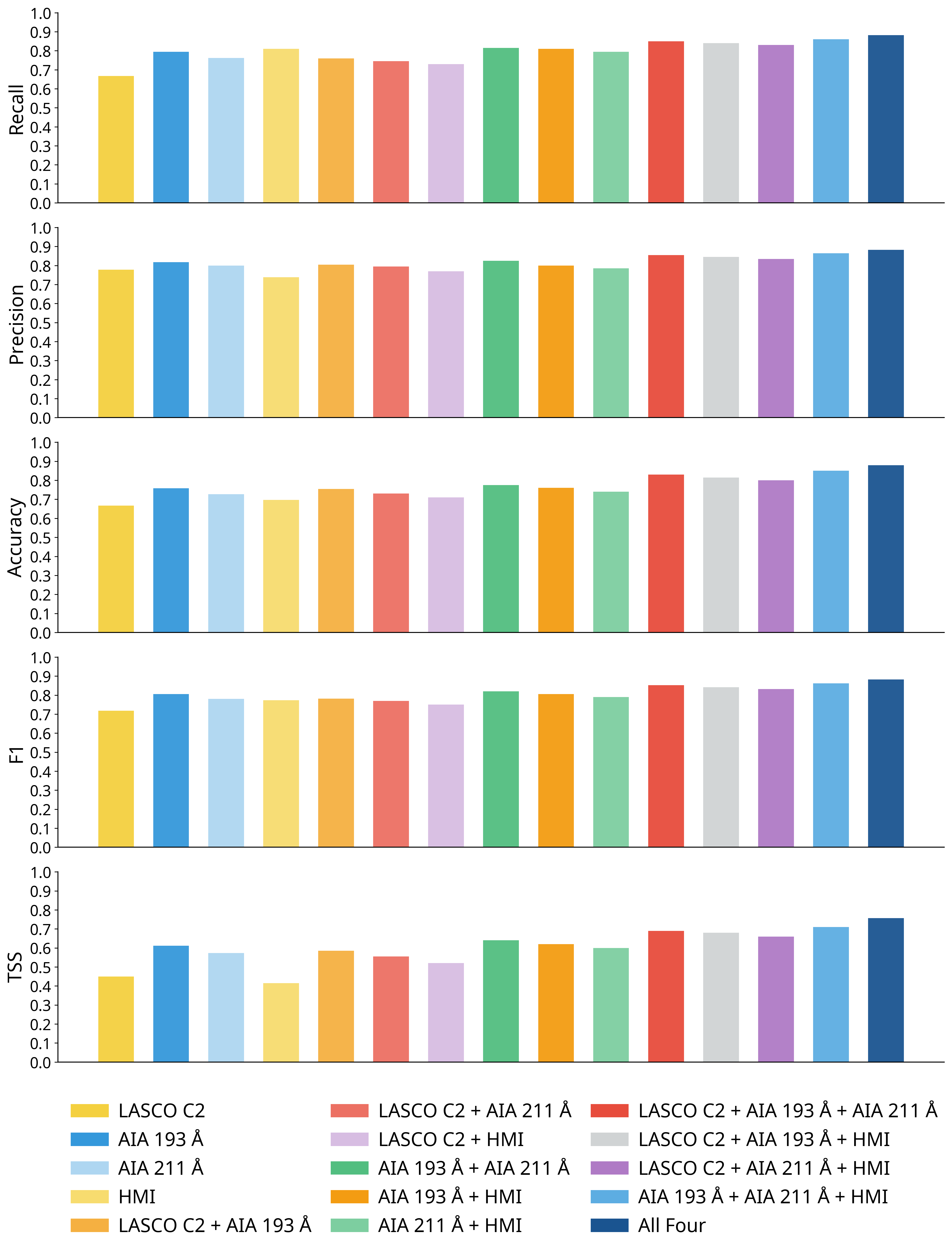}
    \caption{Comparison of the four data types (LASCO C2, AIA 193~\AA{}, AIA 211~\AA{}, HMI)
    and their combinations, each of which is used in turn as input to our fusion model, for deterministic prediction
    based on the 80:20 train-test split described in Section \ref{sec:data}. The model achieves the best
performance when all the four data types are used together.} 
    \label{fig:data_metrics}
\end{figure}

There are four data types
(LASCO C2, AIA 193~\AA{}, AIA 211~\AA{} and HMI)
used as input to our fusion model.
Figure \ref{fig:data_metrics}
compares the effectiveness of the four data types and their combinations,
including
LASCO C2,
AIA 193~\AA{},
AIA 211~\AA{},
HMI, 
LASCO C2 + AIA 193~\AA{},
LASCO C2 + AIA 211~\AA{},
LASCO C2 + HMI,
AIA 193~\AA{} + AIA 211~\AA{},
AIA 193~\AA{} + HMI,
AIA 211~\AA{} + HMI,
LASCO C2 + AIA 193~\AA{} + AIA 211~\AA{},
LASCO C2 + AIA 193~\AA{} + HMI,
LASCO C2 + AIA 211~\AA{} + HMI,
AIA 193~\AA{} + AIA 211~\AA{} + HMI, 
LASCO C2 + AIA 193~\AA{} + AIA 211~\AA{} + HMI.
Each data type alone is used as input 
to our model to produce results.
Then, in turn, each combination is used as input
to the model to produce the results.
It can be seen in Figure \ref{fig:data_metrics} that
combining the four data types together performs the best,
achieving the highest TSS score of 0.757.

\subsection{Model Interpretation}

\begin{figure}[htbp]
    \centering
    \includegraphics[width=\textwidth]{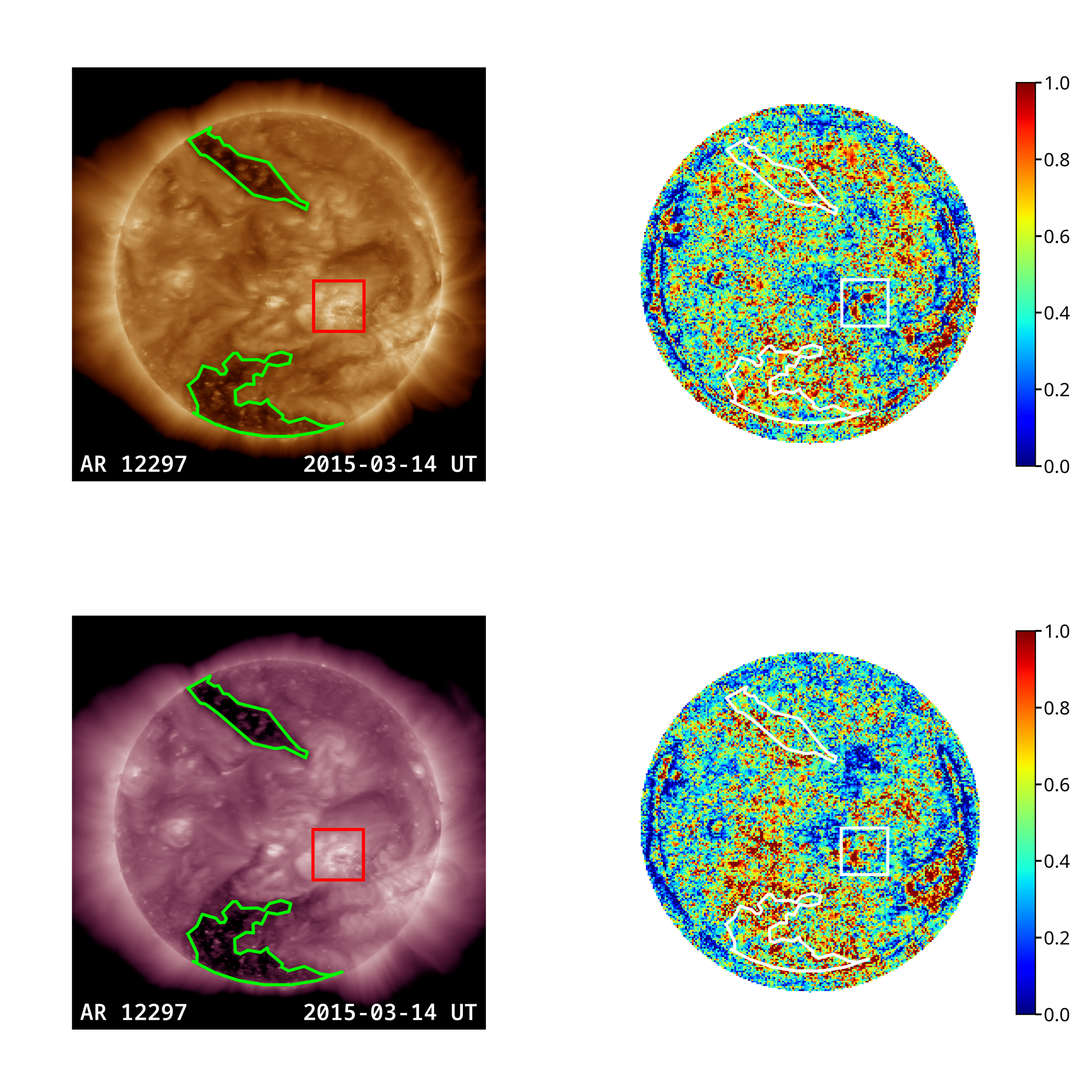}
    \caption{AIA 
    193 \AA{} image (top), AIA 211 \AA{} image (bottom), and their corresponding IG attribution maps 
    for a correctly predicted (true positive)
    CME event that occurred on 14 March 2015 UT.
    The left column shows the AIA images and the right column shows the IG attribution maps.
    Each color bar indicates the normalized attribution intensity, ranging from 0 (blue, low attribution, low importance score) to 1 (red, high attribution, high importance score).
    In each AIA image, the AR 12297, which is the source region of the CME event, is highlighted by a red box, while
    the coronal holes are outlined in green.
    Based on the IG attribution maps, 
    our fusion model assigns high importance scores to the coronal loops in the CME-producing
    AR 12297, 
    paying much attention to the loops,
    and consequently predicts that the
CME event is 
positive/geoeffective.}  
    \label{fig:truepositive}
\end{figure}

To
elucidate the decision-making process of our
fusion model,
we employ 
Integrated Gradients 
\citep[IG;][]{DBLP:conf/icml/SundararajanTY17},
an axiomatic feature attribution method that explains deep learning model predictions by assigning importance scores to input features.
The resulting attribution maps highlight the regions 
in the input images
that most influence the model's predictions.
Based on
Figure \ref{fig:data_metrics},
AIA 193~\AA{} and AIA 211~\AA{} images have a
stronger effect on the performance of our model
than LASCO C2 and HMI images.
Therefore, we use the two AIA images to
explain the model's decision-making process.

Figure \ref{fig:truepositive} 
shows the AIA images and their corresponding
IG attribution maps for 
a correctly 
predicted (true positive) CME event,
denoted $C$,
that occurred on 14 March 2015 UT
(also see Figure \ref{fig:soho_obs}).
In each AIA image, the 
active region (AR) NOAA 12297, which is the source region of the event $C$,
is highlighted by a red box.
This AR 12297 is identified by
(1) locating all solar ARs and
their coordinates
at the appearance time of the event $C$
using the Solar Region Summary
provided by the Space Weather Prediction Center (SWPC),\footnote{
\url{https://www.swpc.noaa.gov/products/solar-region-summary}}
(2) determining which of the ARs located in (1) produces a flare that may possibly be associated with a CME
using the 
Geostationary Operational Environmental Satellite
(GOES) X-ray flare 
catalogs provided by the 
National Centers for Environmental 
Information (NCEI),
and
(3) comparing the location (coordinates) of the AR identified in (2) 
with the location (central position angle) of the CME 
whose onset time recorded in the SOHO LASCO CME Catalog\footnote{
\url{https://cdaw.gsfc.nasa.gov/CME_list/}}
matches, 
within a 2 hr window
\citep{2025ApJ...981...37Z},
the appearance time of the event $C$
to make sure
that the AR identified in (2) is 
the source region of the event $C$.
In addition, we extract
the coronal holes
and their locations (coordinates) in each AIA image 
from the 
Heliophysics Events Knowledgebase 
\citep{2012SoPh..275...67H} using SunPy
\citep{2015CS&D....8a4009S}.
The coronal holes in each AIA image are outlined in green.
Separately, in each IG attribution map,
the AR 12297 is highlighted
by a white box, and the coronal holes are also outlined in white.
In the attribution maps, warmer colors (red) indicate higher attribution values and cooler colors (blue) indicate lower 
attribution values.
It can be seen in 
Figure \ref{fig:truepositive} 
that our fusion model assigns high importance scores to 
the coronal loops in 
    the CME-producing
    AR 12297,
    paying much attention to the loops,
    and consequently predicts that the
CME event $C$ is positive, i.e.,
it will cause a geomagnetic storm. 

\begin{figure}[htbp]
    \centering
    \includegraphics[width=\textwidth]{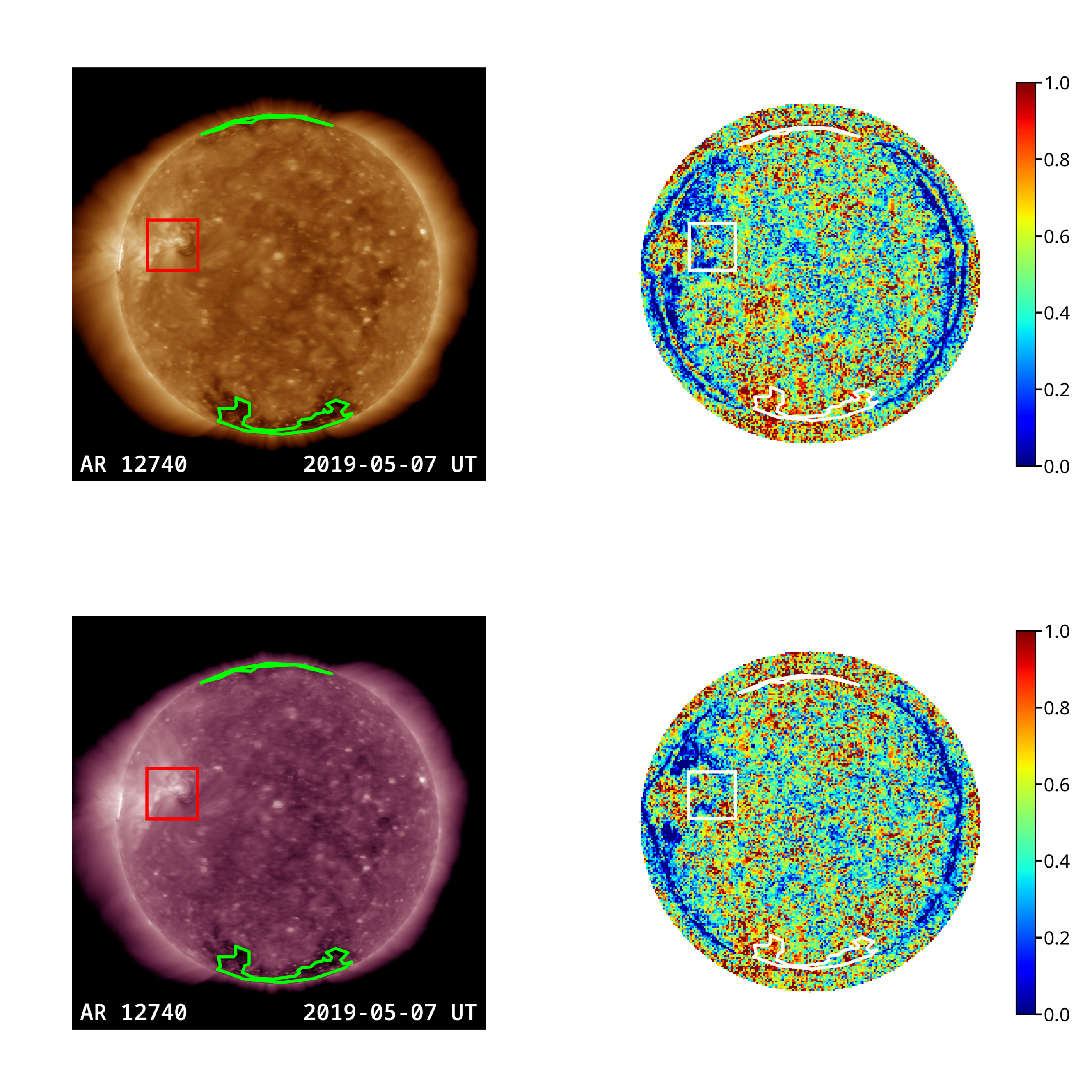}
    \caption{AIA 
    193 \AA{} image (top), 
    AIA 211 \AA{} image (bottom), and their corresponding IG attribution maps 
    for an incorrectly predicted (false negative)
    CME event that occurred on
    7 May 2019 UT.
     In each AIA image, the AR 12740, which is the source region of the CME event, is highlighted by a red box, while
    the coronal holes are outlined in green.
     Based on the IG attribution maps, 
     our fusion model does not pay sufficient attention to the coronal loops in the 
     CME-producing AR 12740.
     Consequently, the model predicts that the CME event is negative, 
     though the CME event is actually 
     positive/geoeffective.}
    \label{fig:falsenegative}
\end{figure}

Figure \ref{fig:falsenegative}
shows the AIA images and 
their corresponding
IG attribution maps for an incorrectly 
predicted
(false negative) CME event
that occurred on 7 May 2019 UT.
In each AIA image, the 
AR 12740, which is the source region of the CME event,
is highlighted by a red box while
the coronal holes are outlined in green.
Separately, in each IG attribution map,
the AR 12740 is highlighted
by a while box, and the coronal holes are also outlined in white.
This AR 12740 and coronal holes 
are located using the same methods as described above.
It can be seen in 
Figure \ref{fig:falsenegative}
that our fusion model does not pay sufficient 
attention to the coronal loops in
the CME-producing AR 12740.
As a result, the model predicts that the CME event is negative, though the CME event is actually positive, i.e., geoeffective.

\subsection{Probabilistic Forecasting Results}

As explained in Figure \ref{fig:fusion_architecture},
our fusion model can be used as a probabilistic forecasting tool,
which outputs a probability that an input CME event
will be geoeffective.
The three component networks (ResNet, EfficientNet, and ViT) can also
be used as probabilistic forecasting tools.
When evaluating these networks for probabilistic forecasting,
we adopted the Brier score \citep{Brier1950} and the Brier skill score \citep{Wilks2010}.
The Brier score, denoted BS, quantifies the accuracy of the probabilistic forecast by computing the mean squared difference between the predicted probabilities and actual binary outcomes, mathematically expressed as:
\begin{equation}
\mathrm{BS} = \frac{1}{N} \sum_{i=1}^{N} (y_i - \hat{y}_i)^2,
\end{equation}
where \(N\) represents the total number of CME events in the test set,
\(y_i\) denotes the actual binary outcome for
the $i$th test event
with 1 indicating that the event is geoeffective
and 0 indicating that the event is non-geoeffective,
and \(\hat{y}_i\) represents the predicted probability for the $i$th test event. 
Brier scores range from 0 to 1, with an optimal value of 0 representing perfect probabilistic accuracy.

The Brier skill score, denoted BSS, provides a normalized measure of model performance, 
calculated as:
\begin{equation}
\mathrm{BSS} = 1 - \frac{\mathrm{BS}}{\frac{1}{N} \sum_{i=1}^{N} (y_i - \bar{y})^2},
\end{equation}
where \(\bar{y}\) = $\frac{1}{N}\sum_{i=1}^{N} y_i$ represents the mean frequency of actual geoeffective events in the test set. 
BSS values range from 
$-\infty$
to 1, with a perfect skill corresponding to BSS = 1.

\begin{figure}
    \centering
    \includegraphics[width=0.6\textwidth]{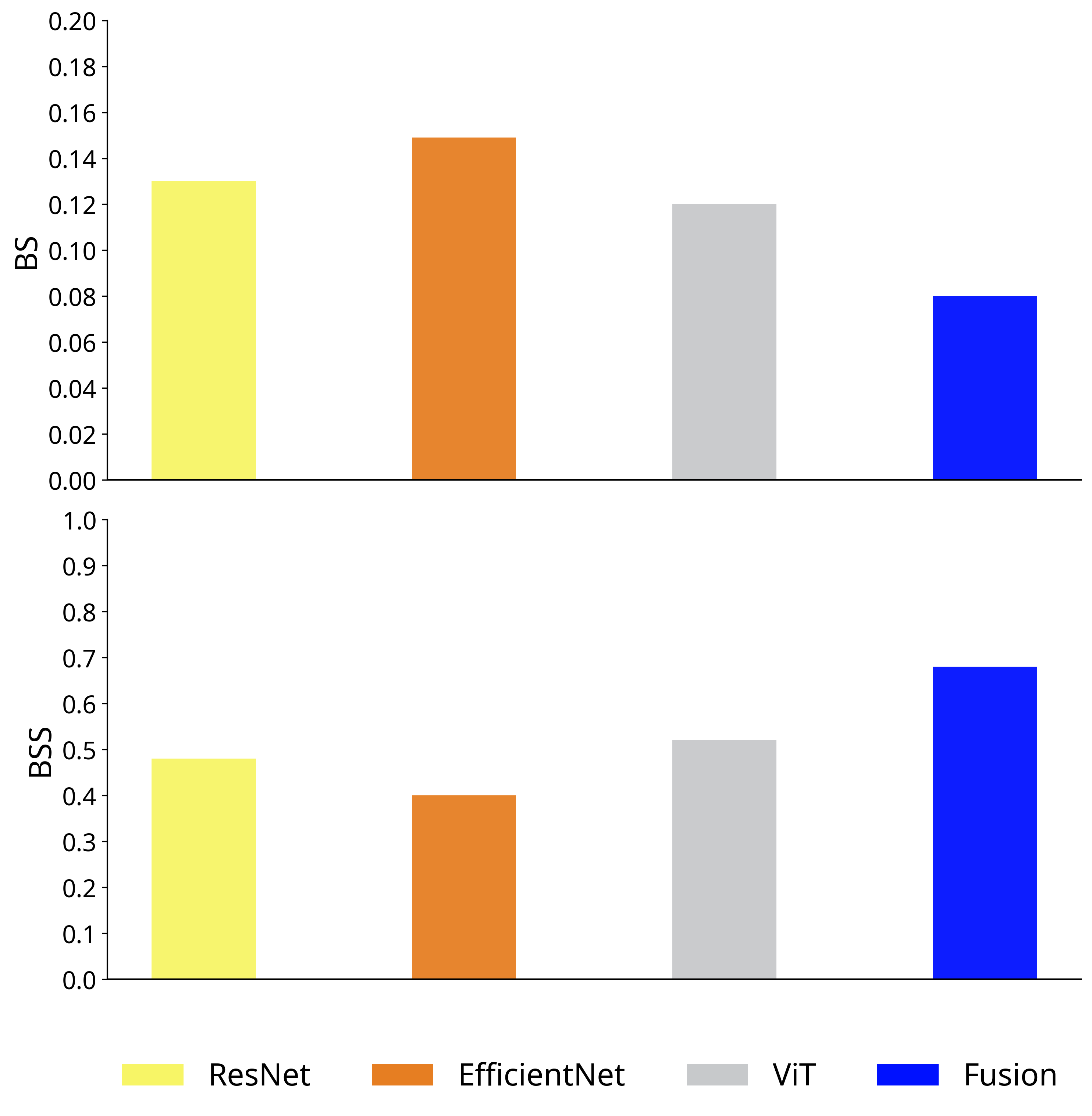}
    \caption{Comparison of our fusion model with its three component networks
    based on the 80:20 train-test split described in Section \ref{sec:data}
    when all the four tools are used for probabilistic forecasting.
    The fusion model achieves the best performance 
    with a BS score of 0.080.} 
    \label{fig:model_bs_bss}
\end{figure}

Figure \ref{fig:model_bs_bss} compares the performance of the
component networks (i.e., the three baseline methods)
and the fusion model for probabilistic forecasting.
Again, ViT performs better than the other two component networks
(ResNet and EfficientNet).
Overall, the fusion model is the best, achieving a
BS score of 0.080,
in reference to
the BS score of 0.094 in \citet{2024SoPh..299..159A}.
To ensure the statistical reliability of our performance assessment,
we also evaluated all models using five-fold cross-validation. 
Across the five folds,
the fusion model
achieved a 
mean BS score
of 0.095,
in reference to
the mean BS score of 
0.107
in
\citet{2024SoPh..299..159A}.
The ResNet baseline
achieved a mean BS score of
0.155, 
the EfficientNet baseline
achieved a mean BS score of
0.179, and
the ViT baseline 
achieved a mean BS score of
0.141. 
Table \ref{tab:cv_probabilistic}
presents the detailed results
 including the mean scores and standard deviations
  of the performance metrics, 
with the best metric values 
highlighted in boldface.
Again, the fusion model
outperforms its component networks.

\begin{table}
\caption{Five-Fold 
Cross-Validation Results of the Four Probabilistic Forecasting Models}
\label{tab:cv_probabilistic}
\begin{tabular}{lcc}
\hline
Model & BS 
& BSS 
\\
\hline
ResNet       & 0.155$\pm$0.012 & 
0.378$\pm$0.048 \\
EfficientNet & 0.179$\pm$0.015 
& 0.282$\pm$0.060 \\
ViT          & 0.141$\pm$0.011 & 
0.435$\pm$0.044 \\
Fusion       & {\bf 0.095}$\pm$0.010 & 
{\bf 0.619}$\pm$0.040 \\
\hline
\end{tabular}
\end{table}

\begin{figure}
    \centering
    \includegraphics[width=0.85\textwidth]{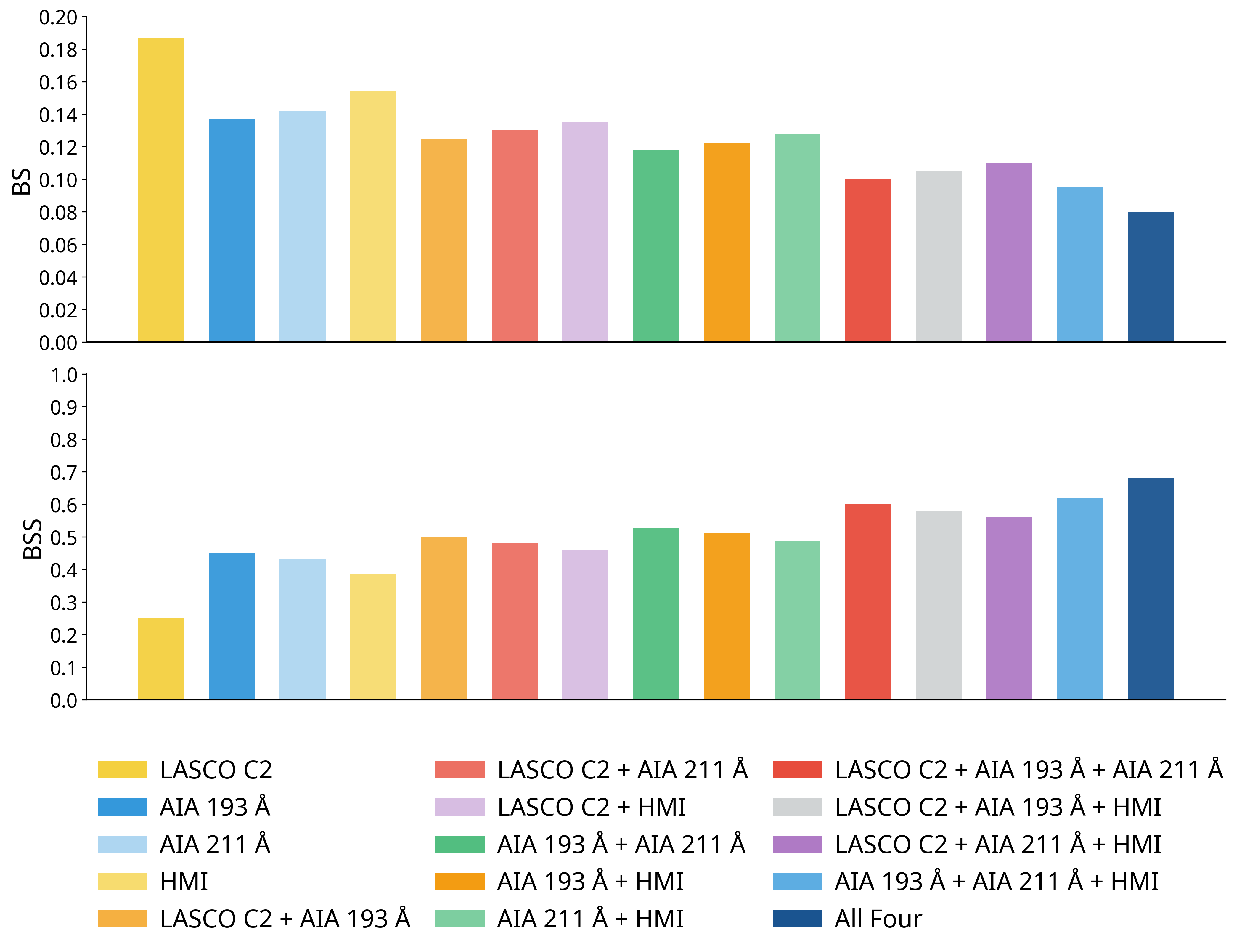}
    \caption{Comparison of the four data types (LASCO C2, AIA 193~\AA{}, AIA 211~\AA{}, HMI)
    and their combinations, each of which is used in turn as input to our fusion model, for probabilistic forecasting
    based on the 80:20 train-test split described in Section \ref{sec:data}. 
    The model achieves the best
performance when all the four data types are used together.} 
    \label{fig:data_bs_bss}
\end{figure}

Figure \ref{fig:data_bs_bss} compares the effectiveness of the four data types
(LASCO C2, AIA 193~\AA{}, AIA 211~\AA{} and HMI)
and their combinations
for probabilistic forecasting.
The results are consistent with those for the deterministic prediction
shown in
Figure \ref{fig:data_metrics} where
combining the four data types together achieves the best performance.

\section{Discussion and Conclusion}
\label{sec:conc}

In this study, we proposed a new fusion model for predicting
geoeffective CMEs, using data from
Solar Cycles (SCs) 24 and 25
to train the model.
Specifically, the model uses four types of solar images (LASCO C2, AIA 193~\AA{}, AIA 211~\AA{} and HMI) and combines three state-of-the-art neural
networks (ResNet, EfficientNet, and ViT)
for feature learning, feature fusion, image classification,
and event prediction.
With five-fold cross validation,
the model achieves a mean true skill statistic (TSS) score of 0.703
when used as a deterministic prediction tool
and a mean Brier score of 0.095 when used as
a probabilistic forecasting tool,
in reference to
the mean TSS score of 0.673 and the
mean Brier score of 0.107 
in \citet{2024SoPh..299..159A},
who used training data collected in SC 23.
Notice that, as in the work of \citet{2024SoPh..299..159A},
we rely on existing tools
\citep[e.g.,][]{Sudar2016,
2018ApJ...855..109L,
2019ApJ...881...15W,
2021SpWea..1902553A,
2021FrASS...8...58D,
2021CosRe..59..268K,
2022FrASS...913345A,
2022A&A...667A.133B,
2023ApJ...954..151G,
2024ApJ...963..121C,
2024ApJ...972...52Y,
2025SpWea..2304165C,
2025ApJ...982...35P}
to predict whether a CME event will arrive at Earth.
If it is predicted that the CME will hit Earth,
then one can use the fusion model developed here
to further predict whether the CME will cause a geomagnetic storm
and/or the probability that the CME will cause such a storm.

Our work uses
AIA images and HMI magnetograms.
In practice,
AIA images can be saturated,
which causes pixels to lose detail, making it difficult to see the underlying structures
\citep{2014ApJ...793L..23S,2019ApJ...882..109G,2022RAA....22f5009Y}.
Saturation occurs when large
solar flares and coronal mass ejections produce intense bursts of EUV light that saturate the detector.
Saturation can be resolved by using de-saturation tools, which reconstruct lost information
\citep{2015A&C....13..117S,2022RAA....22f5009Y}.
In our dataset,
there are 740 AIA 193 \AA{} images
each with 4,096$\times$4,096 pixels;
64.5\% of the images
(i.e., 477 images)
have no saturated pixel. 
Among the remaining 263 images having  saturated pixels, 
33 images contain more than 100 saturated pixels
(out of the  
total 4,096$\times$4,096 = 16,777,216 pixels);
no image contains more than 
1,000 saturated pixels.
There are 646
AIA 211 \AA{} images
each with 4,096$\times$4,096 pixels
(missing images were excluded from
the study as mentioned in Section 
\ref{sec:data});
81.9\% of the images 
(i.e., 529 images)
have no saturated pixel.
Among the remaining 117 images 
having saturated pixels,
31 images contain more than 100 saturated 
pixels, 
12 images contain more than 1,000
saturated pixels, 
and 2 images contain more than
10,000 saturated pixels with one image containing 10,277 saturated pixels and the other image containing 11,506
saturated pixels.
These statistics show that
saturated pixels account for a small portion of the AIA images used in the study.
In addition, 93.5\% of the AIA 193 \AA{}
images have an exposure time  
more than
1.0 second, 
and 91.1\% of the images 
have 
an exposure time  
more than 
2.0 seconds. 
There are 5  AIA 193 \AA{} images 
with an exposure time 
less than 0.2 seconds. 
By contrast, all the 
AIA 211 \AA{} images 
have an exposure time more than 
2.0 seconds.
How saturated and/or short-exposure images influence the prediction of geoeffective CMEs and how to 
explain the influence using IG attribution maps
will be further investigated in future work.

Our experimental results based on
 the 80:20 train-test split and five-fold cross validation
showed that the proposed
fusion model performs better than
all baseline methods.
On the basis of these experimental results,
we conclude that the fusion model is a feasible tool
for predicting geoeffective CMEs.
Because the proposed model is trained by observations collected in the current solar cycle
and the testing SOHO LASCO C2 images as well as the SDO observations can be
obtained in near real-time,
the proposed model has the potential to be an operational
near real-time forecasting system.
\   \\

\noindent
{\scriptsize
{\bf Acknowledgments}
We acknowledge constructive suggestions and comments
from an anonymous referee that have
helped to significantly improve the manuscript. 
We also thank members of the Institute for Space Weather Sciences for fruitful discussions.
The SOHO LASCO CME Catalog is generated and maintained at the CDAW Data Center by NASA and The Catholic University of America in cooperation with the Naval Research Laboratory.
SDO is a NASA mission to observe solar activity and space weather.
SOHO is an international cooperation project between ESA and NASA. 
}

\begin{authorcontribution}
J.W. and H.W. conceived the study.
Z.Y. wrote the manuscript.
All the authors reviewed the manuscript.
\end{authorcontribution}

\begin{fundinginformation}
J.W. and H.W. acknowledge support from 
NSF AGS grants
2228996, 
2149748, 
NSF OAC grants
2504860,
2320147, 
and NASA grants 
80NSSC24K0548,
80NSSC24K0843, 
80NSSC24M0174.
J.J. acknowledges support from NSF AGS grants
2300341, 2149748.
Y.X. acknowledges support from NSF AGS grants
2228996,
2229064, 
and NSF 
RISE-2425602 grant.
V.Y. acknowledges support from
NSF AGS grants
2401229,
2408174,
2300341,
2309939,
NSF AST-2108235 grant, and
NASA 80NSSC24K1914 grant.
\end{fundinginformation}

\begin{dataavailability}
CME events used in this study were compiled from the RC list in
\url{https://izw1.caltech.edu/ACE/ASC/DATA/level3/icmetable2.htm}.
SOHO LASCO C2 images were collected from the
SOHO science archives at
\url{https://soho.nascom.nasa.gov/data/archive/}.
SDO observations were collected from the 
Joint Science Operations Center (JSOC) at
\url{http://jsoc.stanford.edu/}.
\end{dataavailability}

\begin{ethics}
\begin{conflict}
The authors declare no conflict of interest.
\end{conflict}
\end{ethics}

\bibliographystyle{spr-mp-sola}

\begin{thebibliography}{57}
\ifx\bisbn     \undefined \def\bisbn  #1{ISBN #1}\fi
\ifx\binits    \undefined \def\binits#1{#1}\fi
\ifx\bauthor   \undefined \def\bauthor#1{#1}\fi
\ifx\batitle   \undefined \def\batitle#1{#1}\fi
\ifx\bjtitle   \undefined \def\bjtitle#1{\textit{#1}}\fi
\ifx\bvolume   \undefined \def\bvolume#1{\textbf{#1}}\fi
\ifx\byear     \undefined \def\byear#1{#1}\fi
\ifx\bissue    \undefined \def\bissue#1{#1}\fi
\ifx\bfpage    \undefined \def\bfpage#1{#1}\fi
\ifx\blpage    \undefined \def\blpage #1{#1}\fi
\ifx\burl      \undefined \def\burl#1{#1}\fi
\ifx\href      \undefined \def\href#1#2{#2}\fi
\ifx\betal     \undefined \def\betal{et al.}\fi
\ifx\bctitle   \undefined \def\bctitle#1{#1}\fi
\ifx\beditor   \undefined \def\beditor#1{#1}\fi
\ifx\bbtitle   \undefined \def\bbtitle#1{\textit{#1}}\fi
\ifx\bedition  \undefined \def\bedition#1{#1}\fi
\ifx\bseriesno \undefined \def\bseriesno#1{\textbf{#1}}\fi
\ifx\blocation \undefined \def\blocation#1{#1}\fi
\ifx\bsertitle \undefined \def\bsertitle#1{\textit{#1}}\fi
\ifx\bsnm      \undefined \def\bsnm#1{#1}\fi
\ifx\bsuffix   \undefined \def\bsuffix#1{#1}\fi
\ifx\bparticle \undefined \def\bparticle#1{#1}\fi
\ifx\barticle  \undefined \def\barticle#1{}\fi
\ifx\binstitute  \undefined \def\binstitute#1{#1}\fi
\ifx\bpublisher  \undefined \def\bpublisher#1{#1}\fi
\ifx\doiurl    \undefined \def\doiurl#1{\href{#1}{DOI}}\fi
\makeatletter
\def\safeHref#1#2#3{\in@{http}{#2}\ifin@\href{#2}{#3}\else\href{#1#2}{#3}\fi}
\makeatother
\ifx\adsurl    \undefined \def\adsurl#1{\safeHref{https://ui.adsabs.harvard.edu/abs/}{#1}{ADS}}\fi
\ifx\arxivurl  \undefined \def\arxivurl#1{\safeHref{http://arxiv.org/abs/}{#1}{arXiv}}\fi
\ifx\botherref \undefined \def\botherref#1{}\fi
\ifx\url       \undefined \def\url#1{#1}\fi
\ifx\bchapter  \undefined \def\bchapter#1{}\fi
\ifx\bbook     \undefined \def\bbook#1{}\fi
\ifx\bcomment  \undefined \def\bcomment#1{#1}\fi
\ifx\oauthor   \undefined \def\oauthor#1{#1}\fi
\ifx\citeauthoryear \undefined\def \citeauthoryear#1{#1}\fi
\def\endbibitem {}
\ifx\bconflocation  \undefined \def\bconflocation#1{#1} \fi

\bibitem[\protect\citeauthoryear{{Alobaid} et~al.}{2022}]{2022FrASS...913345A}
\begin{barticle}
\bauthor{\bsnm{{Alobaid}}, \binits{K.A.}},
\bauthor{\bsnm{{Abduallah}}, \binits{Y.}},
\bauthor{\bsnm{{Wang}}, \binits{J.T.L.}},
\bauthor{\bsnm{{Wang}}, \binits{H.}},
\bauthor{\bsnm{{Jiang}}, \binits{H.}},
\bauthor{\bsnm{{Xu}}, \binits{Y.}},
\bauthor{\bsnm{{Yurchyshyn}}, \binits{V.}},
\bauthor{\bsnm{{Zhang}}, \binits{H.}},
\bauthor{\bsnm{{Cavus}}, \binits{H.}},
\bauthor{\bsnm{{Jing}}, \binits{J.}}:
\byear{2022},
\batitle{{Predicting CME arrival time through data integration and ensemble learning}}.
\bjtitle{Frontiers in Astronomy and Space Sciences}
\bvolume{9},
\bfpage{1013345}.
\doiurl{https://doi.org/10.3389/fspas.2022.1013345}.
\end{barticle}
\endbibitem

\bibitem[\protect\citeauthoryear{{Alobaid} et~al.}{2023}]{2023ApJ...958L..34A}
\begin{barticle}
\bauthor{\bsnm{{Alobaid}}, \binits{K.A.}},
\bauthor{\bsnm{{Abduallah}}, \binits{Y.}},
\bauthor{\bsnm{{Wang}}, \binits{J.T.L.}},
\bauthor{\bsnm{{Wang}}, \binits{H.}},
\bauthor{\bsnm{{Fan}}, \binits{S.}},
\bauthor{\bsnm{{Li}}, \binits{J.}},
\bauthor{\bsnm{{Cavus}}, \binits{H.}},
\bauthor{\bsnm{{Yurchyshyn}}, \binits{V.}}:
\byear{2023},
\batitle{Estimating coronal mass ejection mass and kinetic energy by fusion of multiple deep-learning models}.
\bjtitle{\apjl}
\bvolume{958},
\bfpage{L34}.
\doiurl{https://doi.org/10.3847/2041-8213/ad0c4a}.
\end{barticle}
\endbibitem

\bibitem[\protect\citeauthoryear{{Alobaid} et~al.}{2024}]{2024SoPh..299..159A}
\begin{barticle}
\bauthor{\bsnm{{Alobaid}}, \binits{K.A.}},
\bauthor{\bsnm{{Wang}}, \binits{J.T.L.}},
\bauthor{\bsnm{{Wang}}, \binits{H.}},
\bauthor{\bsnm{{Jing}}, \binits{J.}},
\bauthor{\bsnm{{Abduallah}}, \binits{Y.}},
\bauthor{\bsnm{{Wang}}, \binits{Z.}},
\bauthor{\bsnm{{Farooki}}, \binits{H.}},
\bauthor{\bsnm{{Cavus}}, \binits{H.}},
\bauthor{\bsnm{{Yurchyshyn}}, \binits{V.}}:
\byear{2024},
\batitle{Prediction of geoeffective {CMEs} using {SOHO} images and deep learning}.
\bjtitle{\solphys}
\bvolume{299},
\bfpage{159}.
\doiurl{https://doi.org/10.1007/s11207-024-02385-w}.
\end{barticle}
\endbibitem

\bibitem[\protect\citeauthoryear{{Amerstorfer} et~al.}{2021}]{2021SpWea..1902553A}
\begin{barticle}
\bauthor{\bsnm{{Amerstorfer}}, \binits{T.}},
\bauthor{\bsnm{{Hinterreiter}}, \binits{J.}},
\bauthor{\bsnm{{Reiss}}, \binits{M.A.}},
\bauthor{\bsnm{{M{\"o}stl}}, \binits{C.}},
\bauthor{\bsnm{{Davies}}, \binits{J.A.}},
\bauthor{\bsnm{{Bailey}}, \binits{R.L.}},
\bauthor{\bsnm{{Weiss}}, \binits{A.J.}},
\bauthor{\bsnm{{Dumbovi{\'c}}}, \binits{M.}},
\bauthor{\bsnm{{Bauer}}, \binits{M.}},
\bauthor{\bsnm{{Amerstorfer}}, \binits{U.V.}},
\bauthor{\bsnm{{Harrison}}, \binits{R.A.}}:
\byear{2021},
\batitle{Evaluation of CME arrival prediction using ensemble modeling based on heliospheric imaging observations}.
\bjtitle{Space Weather}
\bvolume{19},
\bfpage{e02553}.
\doiurl{https://doi.org/10.1029/2020SW002553}.
\end{barticle}
\endbibitem

\bibitem[\protect\citeauthoryear{{Baratashvili} et~al.}{2022}]{2022A&A...667A.133B}
\begin{barticle}
\bauthor{\bsnm{{Baratashvili}}, \binits{T.}},
\bauthor{\bsnm{{Verbeke}}, \binits{C.}},
\bauthor{\bsnm{{Wijsen}}, \binits{N.}},
\bauthor{\bsnm{{Poedts}}, \binits{S.}}:
\byear{2022},
\batitle{{Improving CME evolution and arrival predictions with AMR and grid stretching in Icarus}}.
\bjtitle{\aap}
\bvolume{667},
\bfpage{A133}.
\doiurl{https://doi.org/10.1051/0004-6361/202244111}.
\end{barticle}
\endbibitem

\bibitem[\protect\citeauthoryear{{Besliu-Ionescu} et~al.}{2019}]{2019JASTP.19305036B}
\begin{barticle}
\bauthor{\bsnm{{Besliu-Ionescu}}, \binits{D.}},
\bauthor{\bsnm{{Talpeanu}}, \binits{D.-C.}},
\bauthor{\bsnm{{Mierla}}, \binits{M.}},
\bauthor{\bsnm{{Maris Muntean}}, \binits{G.}}:
\byear{2019},
\batitle{{On the prediction of geoeffectiveness of CMEs during the ascending phase of SC24 using a logistic regression method}}.
\bjtitle{J. Atmos. Solar-Terr. Phys.}
\bvolume{193},
\bfpage{105036}.
\doiurl{https://doi.org/10.1016/j.jastp.2019.04.017}.
\end{barticle}
\endbibitem

\bibitem[\protect\citeauthoryear{{Bloomfield} et~al.}{2012}]{Bloomfield_2012}
\begin{barticle}
\bauthor{\bsnm{{Bloomfield}}, \binits{D.S.}},
\bauthor{\bsnm{{Higgins}}, \binits{P.A.}},
\bauthor{\bsnm{{McAteer}}, \binits{R.T.J.}},
\bauthor{\bsnm{{Gallagher}}, \binits{P.T.}}:
\byear{2012},
\batitle{Toward reliable benchmarking of solar flare forecasting methods}.
\bjtitle{\apjl}
\bvolume{747},
\bfpage{L41}.
\doiurl{https://doi.org/10.1088/2041-8205/747/2/L41}.
\end{barticle}
\endbibitem

\bibitem[\protect\citeauthoryear{Brier}{1950}]{Brier1950}
\begin{barticle}
\bauthor{\bsnm{Brier}, \binits{G.W.}}:
\byear{1950},
\batitle{Verification of forecasts expressed in terms of probability}.
\bjtitle{Monthly Weather Review}
\bvolume{78},
\bfpage{1}.
\doiurl{https://doi.org/10.1175/1520-0493(1950)078<0001:VOFEIT>2.0.CO;2}.
\end{barticle}
\endbibitem

\bibitem[\protect\citeauthoryear{{Brueckner} et~al.}{1995}]{1995SoPh..162..357B}
\begin{barticle}
\bauthor{\bsnm{{Brueckner}}, \binits{G.E.}},
\bauthor{\bsnm{{Howard}}, \binits{R.A.}},
\bauthor{\bsnm{{Koomen}}, \binits{M.J.}},
\bauthor{\bsnm{{Korendyke}}, \binits{C.M.}},
\bauthor{\bsnm{{Michels}}, \binits{D.J.}},
\bauthor{\bsnm{{Moses}}, \binits{J.D.}},
\bauthor{\bsnm{{Socker}}, \binits{D.G.}},
\bauthor{\bsnm{{Dere}}, \binits{K.P.}},
\bauthor{\bsnm{{Lamy}}, \binits{P.L.}},
\bauthor{\bsnm{{Llebaria}}, \binits{A.}},
\bauthor{\bsnm{{Bout}}, \binits{M.V.}},
\bauthor{\bsnm{{Schwenn}}, \binits{R.}},
\bauthor{\bsnm{{Simnett}}, \binits{G.M.}},
\bauthor{\bsnm{{Bedford}}, \binits{D.K.}},
\bauthor{\bsnm{{Eyles}}, \binits{C.J.}}:
\byear{1995},
\batitle{{The Large Angle Spectroscopic Coronagraph (LASCO)}}.
\bjtitle{\solphys}
\bvolume{162},
\bfpage{357}.
\doiurl{https://doi.org/10.1007/BF00733434}.
\end{barticle}
\endbibitem

\bibitem[\protect\citeauthoryear{{Chen} et~al.}{2025}]{2025SpWea..2304165C}
\begin{barticle}
\bauthor{\bsnm{{Chen}}, \binits{H.}},
\bauthor{\bsnm{{Sachdeva}}, \binits{N.}},
\bauthor{\bsnm{{Huang}}, \binits{Z.}},
\bauthor{\bsnm{{Holst}}, \binits{B.}},
\bauthor{\bsnm{{Manchester}}, \binits{W.}},
\bauthor{\bsnm{{Jivani}}, \binits{A.}},
\bauthor{\bsnm{{Zou}}, \binits{S.}},
\bauthor{\bsnm{{Chen}}, \binits{Y.}},
\bauthor{\bsnm{{Huan}}, \binits{X.}},
\bauthor{\bsnm{{Toth}}, \binits{G.}}:
\byear{2025},
\batitle{Decent estimate of CME arrival time from a data-assimilated ensemble in the Alfv{\'e}n Wave Solar atmosphere Model (DECADE-AWSoM)}.
\bjtitle{Space Weather}
\bvolume{23},
\bfpage{2024SW004165}.
\doiurl{https://doi.org/10.1029/2024SW004165}.
\end{barticle}
\endbibitem

\bibitem[\protect\citeauthoryear{{Chierichini} et~al.}{2024}]{2024ApJ...963..121C}
\begin{barticle}
\bauthor{\bsnm{{Chierichini}}, \binits{S.}},
\bauthor{\bsnm{{Liu}}, \binits{J.}},
\bauthor{\bsnm{{Kors{\'o}s}}, \binits{M.B.}},
\bauthor{\bsnm{{Del Moro}}, \binits{D.}},
\bauthor{\bsnm{{Erd{\'e}lyi}}, \binits{R.}}:
\byear{2024},
\batitle{{CME arrival modeling with machine learning}}.
\bjtitle{\apj}
\bvolume{963},
\bfpage{121}.
\doiurl{https://doi.org/10.3847/1538-4357/ad1cee}.
\end{barticle}
\endbibitem

\bibitem[\protect\citeauthoryear{Deng et~al.}{2009}]{Deng2009}
\begin{bchapter}
\bauthor{\bsnm{Deng}, \binits{J.}},
\bauthor{\bsnm{Dong}, \binits{W.}},
\bauthor{\bsnm{Socher}, \binits{R.}},
\bauthor{\bsnm{Li}, \binits{L.-J.}},
\bauthor{\bsnm{Li}, \binits{K.}},
\bauthor{\bsnm{Fei-Fei}, \binits{L.}}:
\byear{2009},
\bctitle{ImageNet: A large-scale hierarchical image database}.
In: \bbtitle{Proceedings of the IEEE Conference on Computer Vision and Pattern Recognition},
\bfpage{248}.
\doiurl{https://doi.org/10.1109/CVPR.2009.5206848}.
\end{bchapter}
\endbibitem

\bibitem[\protect\citeauthoryear{{Dissauer} et~al.}{2018}]{2018ApJ...863..169D}
\begin{barticle}
\bauthor{\bsnm{{Dissauer}}, \binits{K.}},
\bauthor{\bsnm{{Veronig}}, \binits{A.M.}},
\bauthor{\bsnm{{Temmer}}, \binits{M.}},
\bauthor{\bsnm{{Podladchikova}}, \binits{T.}},
\bauthor{\bsnm{{Vanninathan}}, \binits{K.}}:
\byear{2018},
\batitle{Statistics of coronal dimmings associated with coronal mass ejections. I. Characteristic dimming properties and flare association}.
\bjtitle{\apj}
\bvolume{863},
\bfpage{169}.
\doiurl{https://doi.org/10.3847/1538-4357/aad3c6}.
\end{barticle}
\endbibitem

\bibitem[\protect\citeauthoryear{{Dissauer} et~al.}{2019}]{2019ApJ...874..123D}
\begin{barticle}
\bauthor{\bsnm{{Dissauer}}, \binits{K.}},
\bauthor{\bsnm{{Veronig}}, \binits{A.M.}},
\bauthor{\bsnm{{Temmer}}, \binits{M.}},
\bauthor{\bsnm{{Podladchikova}}, \binits{T.}}:
\byear{2019},
\batitle{Statistics of coronal dimmings associated with coronal mass ejections. II. Relationship between coronal dimmings and their associated CMEs}.
\bjtitle{\apj}
\bvolume{874},
\bfpage{123}.
\doiurl{https://doi.org/10.3847/1538-4357/ab0962}.
\end{barticle}
\endbibitem

\bibitem[\protect\citeauthoryear{{Domingo}, {Fleck}, and {Poland}}{1995}]{1995SoPh..162....1D}
\begin{barticle}
\bauthor{\bsnm{{Domingo}}, \binits{V.}},
\bauthor{\bsnm{{Fleck}}, \binits{B.}},
\bauthor{\bsnm{{Poland}}, \binits{A.I.}}:
\byear{1995},
\batitle{{The SOHO mission: An overview}}.
\bjtitle{\solphys}
\bvolume{162},
\bfpage{1}.
\doiurl{https://doi.org/10.1007/BF00733425}.
\end{barticle}
\endbibitem

\bibitem[\protect\citeauthoryear{Dosovitskiy et~al.}{2021}]{DBLP:conf/iclr/DosovitskiyB0WZ21}
\begin{bchapter}
\bauthor{\bsnm{Dosovitskiy}, \binits{A.}},
\bauthor{\bsnm{Beyer}, \binits{L.}},
\bauthor{\bsnm{Kolesnikov}, \binits{A.}},
\bauthor{\bsnm{Weissenborn}, \binits{D.}},
\bauthor{\bsnm{Zhai}, \binits{X.}},
\bauthor{\bsnm{Unterthiner}, \binits{T.}},
\bauthor{\bsnm{Dehghani}, \binits{M.}},
\bauthor{\bsnm{Minderer}, \binits{M.}},
\bauthor{\bsnm{Heigold}, \binits{G.}},
\bauthor{\bsnm{Gelly}, \binits{S.}},
\bauthor{\bsnm{Uszkoreit}, \binits{J.}},
\bauthor{\bsnm{Houlsby}, \binits{N.}}:
\byear{2021},
\bctitle{An image is worth 16x16 words: Transformers for image recognition at scale}.
In: \bbtitle{Proceedings of the 9th International Conference on Learning Representations}.
\end{bchapter}
\endbibitem

\bibitem[\protect\citeauthoryear{{Dumbovi{\'c}} et~al.}{2021}]{2021FrASS...8...58D}
\begin{barticle}
\bauthor{\bsnm{{Dumbovi{\'c}}}, \binits{M.}},
\bauthor{\bsnm{{{\v{C}}alogovi{\'c}}}, \binits{J.}},
\bauthor{\bsnm{{Martini{\'c}}}, \binits{K.}},
\bauthor{\bsnm{{Vr{\v{s}}nak}}, \binits{B.}},
\bauthor{\bsnm{{Sudar}}, \binits{D.}},
\bauthor{\bsnm{{Temmer}}, \binits{M.}},
\bauthor{\bsnm{{Veronig}}, \binits{A.}}:
\byear{2021},
\batitle{{Drag-based model (DBM) tools for forecast of coronal mass ejection arrival time and speed}}.
\bjtitle{Front. Astron. Space Sci.}
\bvolume{8},
\bfpage{58}.
\doiurl{https://doi.org/10.3389/fspas.2021.639986}.
\end{barticle}
\endbibitem

\bibitem[\protect\citeauthoryear{{Falconer}, {Moore}, and {Gary}}{2006}]{2006ApJ...644.1258F}
\begin{barticle}
\bauthor{\bsnm{{Falconer}}, \binits{D.A.}},
\bauthor{\bsnm{{Moore}}, \binits{R.L.}},
\bauthor{\bsnm{{Gary}}, \binits{G.A.}}:
\byear{2006},
\batitle{{Magnetic causes of solar coronal mass ejections: Dominance of the free magnetic energy over the magnetic twist alone}}.
\bjtitle{\apj}
\bvolume{644},
\bfpage{1258}.
\doiurl{https://doi.org/10.1086/503699}.
\end{barticle}
\endbibitem

\bibitem[\protect\citeauthoryear{{Fu} et~al.}{2021}]{2021RemS...13.1738F}
\begin{barticle}
\bauthor{\bsnm{{Fu}}, \binits{H.}},
\bauthor{\bsnm{{Zheng}}, \binits{Y.}},
\bauthor{\bsnm{{Ye}}, \binits{Y.}},
\bauthor{\bsnm{{Feng}}, \binits{X.}},
\bauthor{\bsnm{{Liu}}, \binits{C.}},
\bauthor{\bsnm{{Ma}}, \binits{H.}}:
\byear{2021},
\batitle{{Joint geoeffectiveness and arrival time prediction of CMEs by a unified deep learning framework}}.
\bjtitle{Remote Sens.}
\bvolume{13},
\bfpage{1738}.
\doiurl{https://doi.org/10.3390/rs13091738}.
\end{barticle}
\endbibitem

\bibitem[\protect\citeauthoryear{Gonzalez et~al.}{1994}]{Gonzalez1994}
\begin{barticle}
\bauthor{\bsnm{Gonzalez}, \binits{W.D.}},
\bauthor{\bsnm{Joselyn}, \binits{J.A.}},
\bauthor{\bsnm{Kamide}, \binits{Y.}},
\bauthor{\bsnm{Kroehl}, \binits{H.W.}},
\bauthor{\bsnm{Rostoker}, \binits{G.}},
\bauthor{\bsnm{Tsurutani}, \binits{B.T.}},
\bauthor{\bsnm{Vasyliunas}, \binits{V.M.}}:
\byear{1994},
\batitle{What is a geomagnetic storm?}
\bjtitle{\jgr}
\bvolume{99},
\bfpage{5771}.
\doiurl{https://doi.org/10.1029/93JA02867}.
\end{barticle}
\endbibitem

\bibitem[\protect\citeauthoryear{{Gopalswamy}}{2009}]{2009EP&S...61..595G}
\begin{barticle}
\bauthor{\bsnm{{Gopalswamy}}, \binits{N.}}:
\byear{2009},
\batitle{Halo coronal mass ejections and geomagnetic storms}.
\bjtitle{Earth, Planets and Space}
\bvolume{61},
\bfpage{595}.
\doiurl{https://doi.org/10.1186/BF03352930}.
\end{barticle}
\endbibitem

\bibitem[\protect\citeauthoryear{{Gopalswamy} et~al.}{2009}]{2009JGRA..114.0A22G}
\begin{barticle}
\bauthor{\bsnm{{Gopalswamy}}, \binits{N.}},
\bauthor{\bsnm{{M{\"a}kel{\"a}}}, \binits{P.}},
\bauthor{\bsnm{{Xie}}, \binits{H.}},
\bauthor{\bsnm{{Akiyama}}, \binits{S.}},
\bauthor{\bsnm{{Yashiro}}, \binits{S.}}:
\byear{2009},
\batitle{{CME interactions with coronal holes and their interplanetary consequences}}.
\bjtitle{Journal of Geophysical Research (Space Physics)}
\bvolume{114},
\bfpage{A00A22}.
\doiurl{https://doi.org/10.1029/2008JA013686}.
\end{barticle}
\endbibitem

\bibitem[\protect\citeauthoryear{{Guastavino} et~al.}{2019}]{2019ApJ...882..109G}
\begin{barticle}
\bauthor{\bsnm{{Guastavino}}, \binits{S.}},
\bauthor{\bsnm{{Piana}}, \binits{M.}},
\bauthor{\bsnm{{Massone}}, \binits{A.M.}},
\bauthor{\bsnm{{Schwartz}}, \binits{R.}},
\bauthor{\bsnm{{Benvenuto}}, \binits{F.}}:
\byear{2019},
\batitle{{Desaturating SDO/AIA observations of solar flaring storms}}.
\bjtitle{\apj}
\bvolume{882},
\bfpage{109}.
\doiurl{https://doi.org/10.3847/1538-4357/ab35d8}.
\end{barticle}
\endbibitem

\bibitem[\protect\citeauthoryear{{Guastavino} et~al.}{2023}]{2023ApJ...954..151G}
\begin{barticle}
\bauthor{\bsnm{{Guastavino}}, \binits{S.}},
\bauthor{\bsnm{{Candiani}}, \binits{V.}},
\bauthor{\bsnm{{Bemporad}}, \binits{A.}},
\bauthor{\bsnm{{Marchetti}}, \binits{F.}},
\bauthor{\bsnm{{Benvenuto}}, \binits{F.}},
\bauthor{\bsnm{{Massone}}, \binits{A.M.}},
\bauthor{\bsnm{{Mancuso}}, \binits{S.}},
\bauthor{\bsnm{{Susino}}, \binits{R.}},
\bauthor{\bsnm{{Telloni}}, \binits{D.}},
\bauthor{\bsnm{{Fineschi}}, \binits{S.}},
\bauthor{\bsnm{{Piana}}, \binits{M.}}:
\byear{2023},
\batitle{{Physics-driven machine learning for the prediction of coronal mass ejections' travel times}}.
\bjtitle{\apj}
\bvolume{954},
\bfpage{151}.
\doiurl{https://doi.org/10.3847/1538-4357/ace62d}.
\end{barticle}
\endbibitem

\bibitem[\protect\citeauthoryear{{Guastavino} et~al.}{2024}]{Guastavino_2024}
\begin{barticle}
\bauthor{\bsnm{{Guastavino}}, \binits{S.}},
\bauthor{\bsnm{{Bahamazava}}, \binits{K.}},
\bauthor{\bsnm{{Perracchione}}, \binits{E.}},
\bauthor{\bsnm{{Camattari}}, \binits{F.}},
\bauthor{\bsnm{{Audone}}, \binits{G.}},
\bauthor{\bsnm{{Telloni}}, \binits{D.}},
\bauthor{\bsnm{{Susino}}, \binits{R.}},
\bauthor{\bsnm{{Nicolini}}, \binits{G.}},
\bauthor{\bsnm{{Fineschi}}, \binits{S.}},
\bauthor{\bsnm{{Piana}}, \binits{M.}},
\bauthor{\bsnm{{Massone}}, \binits{A.M.}}:
\byear{2024},
\batitle{Forecasting geoffective events from solar wind data and evaluating the most predictive features through machine learning approaches}.
\bjtitle{\apj}
\bvolume{971},
\bfpage{94}.
\doiurl{https://doi.org/10.3847/1538-4357/ad5b57}.
\end{barticle}
\endbibitem

\bibitem[\protect\citeauthoryear{He et~al.}{2016}]{He2016}
\begin{bchapter}
\bauthor{\bsnm{He}, \binits{K.}},
\bauthor{\bsnm{Zhang}, \binits{X.}},
\bauthor{\bsnm{Ren}, \binits{S.}},
\bauthor{\bsnm{Sun}, \binits{J.}}:
\byear{2016},
\bctitle{Deep residual learning for image recognition}.
In: \bbtitle{Proceedings of the IEEE Conference on Computer Vision and Pattern Recognition},
\bfpage{770}.
\doiurl{https://doi.org/10.1109/CVPR.2016.90}.
\end{bchapter}
\endbibitem

\bibitem[\protect\citeauthoryear{Hu, Camporeale, and Swiger}{2023}]{Hu_2023}
\begin{barticle}
\bauthor{\bsnm{Hu}, \binits{A.}},
\bauthor{\bsnm{Camporeale}, \binits{E.}},
\bauthor{\bsnm{Swiger}, \binits{B.}}:
\byear{2023},
\batitle{Multi‐hour‐ahead Dst index prediction using multi‐fidelity boosted neural networks}.
\bjtitle{Space Weather}
\bvolume{21}.
\doiurl{https://doi.org/10.1029/2022sw003286}.
\end{barticle}
\endbibitem

\bibitem[\protect\citeauthoryear{{Hurlburt} et~al.}{2012}]{2012SoPh..275...67H}
\begin{barticle}
\bauthor{\bsnm{{Hurlburt}}, \binits{N.}},
\bauthor{\bsnm{{Cheung}}, \binits{M.}},
\bauthor{\bsnm{{Schrijver}}, \binits{C.}},
\bauthor{\bsnm{{Chang}}, \binits{L.}},
\bauthor{\bsnm{{Freeland}}, \binits{S.}},
\bauthor{\bsnm{{Green}}, \binits{S.}},
\bauthor{\bsnm{{Heck}}, \binits{C.}},
\bauthor{\bsnm{{Jaffey}}, \binits{A.}},
\bauthor{\bsnm{{Kobashi}}, \binits{A.}},
\bauthor{\bsnm{{Schiff}}, \binits{D.}},
\bauthor{\bsnm{{Serafin}}, \binits{J.}},
\bauthor{\bsnm{{Seguin}}, \binits{R.}},
\bauthor{\bsnm{{Slater}}, \binits{G.}},
\bauthor{\bsnm{{Somani}}, \binits{A.}},
\bauthor{\bsnm{{Timmons}}, \binits{R.}}:
\byear{2012},
\batitle{{Heliophysics Event Knowledgebase} for the {Solar Dynamics Observatory (SDO)} and beyond}.
\bjtitle{\solphys}
\bvolume{275},
\bfpage{67}.
\doiurl{https://doi.org/10.1007/s11207-010-9624-2}.
\end{barticle}
\endbibitem

\bibitem[\protect\citeauthoryear{{Kaportseva} and {Shugay}}{2021}]{2021CosRe..59..268K}
\begin{barticle}
\bauthor{\bsnm{{Kaportseva}}, \binits{K.B.}},
\bauthor{\bsnm{{Shugay}}, \binits{Y.S.}}:
\byear{2021},
\batitle{{Use of the DBM model to the predict of arrival of coronal mass ejections to the Earth}}.
\bjtitle{Cosmic Res.}
\bvolume{59},
\bfpage{268}.
\doiurl{https://doi.org/10.1134/S001095252104002X}.
\end{barticle}
\endbibitem

\bibitem[\protect\citeauthoryear{{Lemen} et~al.}{2012}]{2012SoPh..275...17L}
\begin{barticle}
\bauthor{\bsnm{{Lemen}}, \binits{J.R.}},
\bauthor{\bsnm{{Title}}, \binits{A.M.}},
\bauthor{\bsnm{{Akin}}, \binits{D.J.}},
\bauthor{\bsnm{{Boerner}}, \binits{P.F.}},
\bauthor{\bsnm{{Chou}}, \binits{C.}},
\bauthor{\bsnm{{Drake}}, \binits{J.F.}},
\bauthor{\bsnm{{Duncan}}, \binits{D.W.}},
\bauthor{\bsnm{{Edwards}}, \binits{C.G.}},
\bauthor{\bsnm{{Friedlaender}}, \binits{F.M.}},
\bauthor{\bsnm{{Heyman}}, \binits{G.F.}},
\bauthor{\bsnm{{Hurlburt}}, \binits{N.E.}},
\bauthor{\bsnm{{Katz}}, \binits{N.L.}},
\bauthor{\bsnm{{Kushner}}, \binits{G.D.}},
\bauthor{\bsnm{{Levay}}, \binits{M.}},
\bauthor{\bsnm{{Lindgren}}, \binits{R.W.}},
\bauthor{\bsnm{{Mathur}}, \binits{D.P.}},
\bauthor{\bsnm{{McFeaters}}, \binits{E.L.}},
\bauthor{\bsnm{{Mitchell}}, \binits{S.}},
\bauthor{\bsnm{{Rehse}}, \binits{R.A.}},
\bauthor{\bsnm{{Schrijver}}, \binits{C.J.}},
\bauthor{\bsnm{{Springer}}, \binits{L.A.}},
\bauthor{\bsnm{{Stern}}, \binits{R.A.}},
\bauthor{\bsnm{{Tarbell}}, \binits{T.D.}},
\bauthor{\bsnm{{Wuelser}}, \binits{J.-P.}},
\bauthor{\bsnm{{Wolfson}}, \binits{C.J.}},
\bauthor{\bsnm{{Yanari}}, \binits{C.}},
\bauthor{\bsnm{{Bookbinder}}, \binits{J.A.}},
\bauthor{\bsnm{{Cheimets}}, \binits{P.N.}},
\bauthor{\bsnm{{Caldwell}}, \binits{D.}},
\bauthor{\bsnm{{Deluca}}, \binits{E.E.}},
\bauthor{\bsnm{{Gates}}, \binits{R.}},
\bauthor{\bsnm{{Golub}}, \binits{L.}},
\bauthor{\bsnm{{Park}}, \binits{S.}},
\bauthor{\bsnm{{Podgorski}}, \binits{W.A.}},
\bauthor{\bsnm{{Bush}}, \binits{R.I.}},
\bauthor{\bsnm{{Scherrer}}, \binits{P.H.}},
\bauthor{\bsnm{{Gummin}}, \binits{M.A.}},
\bauthor{\bsnm{{Smith}}, \binits{P.}},
\bauthor{\bsnm{{Auker}}, \binits{G.}},
\bauthor{\bsnm{{Jerram}}, \binits{P.}},
\bauthor{\bsnm{{Pool}}, \binits{P.}},
\bauthor{\bsnm{{Soufli}}, \binits{R.}},
\bauthor{\bsnm{{Windt}}, \binits{D.L.}},
\bauthor{\bsnm{{Beardsley}}, \binits{S.}},
\bauthor{\bsnm{{Clapp}}, \binits{M.}},
\bauthor{\bsnm{{Lang}}, \binits{J.}},
\bauthor{\bsnm{{Waltham}}, \binits{N.}}:
\byear{2012},
\batitle{{The Atmospheric Imaging Assembly (AIA) on the Solar Dynamics Observatory (SDO)}}.
\bjtitle{\solphys}
\bvolume{275},
\bfpage{17}.
\doiurl{https://doi.org/10.1007/s11207-011-9776-8}.
\end{barticle}
\endbibitem

\bibitem[\protect\citeauthoryear{{Liu} et~al.}{2018}]{2018ApJ...855..109L}
\begin{barticle}
\bauthor{\bsnm{{Liu}}, \binits{J.}},
\bauthor{\bsnm{{Ye}}, \binits{Y.}},
\bauthor{\bsnm{{Shen}}, \binits{C.}},
\bauthor{\bsnm{{Wang}}, \binits{Y.}},
\bauthor{\bsnm{{Erd{\'e}lyi}}, \binits{R.}}:
\byear{2018},
\batitle{A new tool for {CME} arrival time prediction using machine learning algorithms: {CAT-PUMA}}.
\bjtitle{\apj}
\bvolume{855},
\bfpage{109}.
\doiurl{https://doi.org/10.3847/1538-4357/aaae69}.
\end{barticle}
\endbibitem

\bibitem[\protect\citeauthoryear{Liu et~al.}{2024}]{Liu2024}
\begin{barticle}
\bauthor{\bsnm{Liu}, \binits{J.}},
\bauthor{\bsnm{Shen}, \binits{C.}},
\bauthor{\bsnm{Wang}, \binits{Y.}},
\bauthor{\bsnm{Xu}, \binits{M.}},
\bauthor{\bsnm{Chi}, \binits{Y.}},
\bauthor{\bsnm{Zhong}, \binits{Z.}},
\bauthor{\bsnm{Mao}, \binits{D.}},
\bauthor{\bsnm{Zhang}, \binits{Z.}},
\bauthor{\bsnm{Wang}, \binits{C.}},
\bauthor{\bsnm{Liu}, \binits{J.}},
\bauthor{\bsnm{Wang}, \binits{Y.}}:
\byear{2024},
\batitle{Forecasting the {Dst} index with temporal convolutional network and integrated gradients}.
\bjtitle{\solphys}
\bvolume{299},
\bfpage{98}.
\doiurl{https://doi.org/10.1007/s11207-024-02340-9}.
\end{barticle}
\endbibitem

\bibitem[\protect\citeauthoryear{{Liu}, {Webb}, and {Zhao}}{2006}]{2006ApJ...646.1335L}
\begin{barticle}
\bauthor{\bsnm{{Liu}}, \binits{Y.}},
\bauthor{\bsnm{{Webb}}, \binits{D.F.}},
\bauthor{\bsnm{{Zhao}}, \binits{X.P.}}:
\byear{2006},
\batitle{Magnetic structures of solar active regions, full-halo coronal mass ejections, and geomagnetic storms}.
\bjtitle{\apj}
\bvolume{646},
\bfpage{1335}.
\doiurl{https://doi.org/10.1086/505036}.
\end{barticle}
\endbibitem

\bibitem[\protect\citeauthoryear{Mayaud}{1980}]{doi:https://doi.org/10.1002/9781118663837.ch2}
\begin{bchapter}
\bauthor{\bsnm{Mayaud}, \binits{P.N.}}:
\byear{1980},
\bctitle{What is a geomagnetic index?}
In: \bbtitle{Derivation, Meaning, and Use of Geomagnetic Indices},
\bpublisher{American Geophysical Union}.
\bcomment{Chap. 2}.
\doiurl{https://doi.org/10.1002/9781118663837.ch2}.
\end{bchapter}
\endbibitem

\bibitem[\protect\citeauthoryear{Pan and Yang}{2010}]{DBLP:journals/tkde/PanY10}
\begin{barticle}
\bauthor{\bsnm{Pan}, \binits{S.J.}},
\bauthor{\bsnm{Yang}, \binits{Q.}}:
\byear{2010},
\batitle{A survey on transfer learning}.
\bjtitle{{IEEE} Trans. Knowl. Data Eng.}
\bvolume{22},
\bfpage{1345}.
\doiurl{https://doi.org/10.1109/TKDE.2009.191}.
\end{barticle}
\endbibitem

\bibitem[\protect\citeauthoryear{{Pattnaik}, {Micha{\l}ek}, and {Ravishankar}}{2025}]{2025ApJ...982...35P}
\begin{barticle}
\bauthor{\bsnm{{Pattnaik}}, \binits{A.}},
\bauthor{\bsnm{{Micha{\l}ek}}, \binits{G.}},
\bauthor{\bsnm{{Ravishankar}}, \binits{A.}}:
\byear{2025},
\batitle{{Estimation of transit time of CMEs from the Sun to Earth}}.
\bjtitle{\apj}
\bvolume{982},
\bfpage{35}.
\doiurl{https://doi.org/10.3847/1538-4357/adb84b}.
\end{barticle}
\endbibitem

\bibitem[\protect\citeauthoryear{{Pesnell}, {Thompson}, and {Chamberlin}}{2012}]{2012SoPh..275....3P}
\begin{barticle}
\bauthor{\bsnm{{Pesnell}}, \binits{W.D.}},
\bauthor{\bsnm{{Thompson}}, \binits{B.J.}},
\bauthor{\bsnm{{Chamberlin}}, \binits{P.C.}}:
\byear{2012},
\batitle{{The Solar Dynamics Observatory (SDO)}}.
\bjtitle{\solphys}
\bvolume{275},
\bfpage{3}.
\doiurl{https://doi.org/10.1007/s11207-011-9841-3}.
\end{barticle}
\endbibitem

\bibitem[\protect\citeauthoryear{{Pricopi} et~al.}{2022}]{2022ApJ...934..176P}
\begin{barticle}
\bauthor{\bsnm{{Pricopi}}, \binits{A.-C.}},
\bauthor{\bsnm{{Paraschiv}}, \binits{A.R.}},
\bauthor{\bsnm{{Besliu-Ionescu}}, \binits{D.}},
\bauthor{\bsnm{{Marginean}}, \binits{A.-N.}}:
\byear{2022},
\batitle{{Predicting the geoeffectiveness of CMEs using machine learning}}.
\bjtitle{\apj}
\bvolume{934},
\bfpage{176}.
\doiurl{https://doi.org/10.3847/1538-4357/ac7962}.
\end{barticle}
\endbibitem

\bibitem[\protect\citeauthoryear{{Richardson} and {Cane}}{2010}]{2010SoPh..264..189R}
\begin{barticle}
\bauthor{\bsnm{{Richardson}}, \binits{I.G.}},
\bauthor{\bsnm{{Cane}}, \binits{H.V.}}:
\byear{2010},
\batitle{Near-{Earth} interplanetary coronal mass ejections during solar cycle 23 (1996 - 2009): Catalog and summary of properties}.
\bjtitle{\solphys}
\bvolume{264},
\bfpage{189}.
\doiurl{https://doi.org/10.1007/s11207-010-9568-6}.
\end{barticle}
\endbibitem

\bibitem[\protect\citeauthoryear{{Scherrer} et~al.}{2012}]{2012SoPh..275..207S}
\begin{barticle}
\bauthor{\bsnm{{Scherrer}}, \binits{P.H.}},
\bauthor{\bsnm{{Schou}}, \binits{J.}},
\bauthor{\bsnm{{Bush}}, \binits{R.I.}},
\bauthor{\bsnm{{Kosovichev}}, \binits{A.G.}},
\bauthor{\bsnm{{Bogart}}, \binits{R.S.}},
\bauthor{\bsnm{{Hoeksema}}, \binits{J.T.}},
\bauthor{\bsnm{{Liu}}, \binits{Y.}},
\bauthor{\bsnm{{Duvall}}, \binits{T.L.}},
\bauthor{\bsnm{{Zhao}}, \binits{J.}},
\bauthor{\bsnm{{Title}}, \binits{A.M.}},
\bauthor{\bsnm{{Schrijver}}, \binits{C.J.}},
\bauthor{\bsnm{{Tarbell}}, \binits{T.D.}},
\bauthor{\bsnm{{Tomczyk}}, \binits{S.}}:
\byear{2012},
\batitle{{The Helioseismic and Magnetic Imager (HMI) investigation for the Solar Dynamics Observatory (SDO)}}.
\bjtitle{\solphys}
\bvolume{275},
\bfpage{207}.
\doiurl{https://doi.org/10.1007/s11207-011-9834-2}.
\end{barticle}
\endbibitem

\bibitem[\protect\citeauthoryear{{Schwartz}, {Torre}, and {Piana}}{2014}]{2014ApJ...793L..23S}
\begin{barticle}
\bauthor{\bsnm{{Schwartz}}, \binits{R.A.}},
\bauthor{\bsnm{{Torre}}, \binits{G.}},
\bauthor{\bsnm{{Piana}}, \binits{M.}}:
\byear{2014},
\batitle{Systematic de-saturation of images from the Atmospheric Imaging Assembly in the Solar Dynamics Observatory}.
\bjtitle{\apjl}
\bvolume{793},
\bfpage{L23}.
\doiurl{https://doi.org/10.1088/2041-8205/793/2/L23}.
\end{barticle}
\endbibitem

\bibitem[\protect\citeauthoryear{{Schwartz} et~al.}{2015}]{2015A&C....13..117S}
\begin{barticle}
\bauthor{\bsnm{{Schwartz}}, \binits{R.A.}},
\bauthor{\bsnm{{Torre}}, \binits{G.}},
\bauthor{\bsnm{{Massone}}, \binits{A.M.}},
\bauthor{\bsnm{{Piana}}, \binits{M.}}:
\byear{2015},
\batitle{{DESAT: A Solar SoftWare tool for image de-saturation in the Atmospheric Image Assembly onboard the Solar Dynamics Observatory}}.
\bjtitle{Astronomy and Computing}
\bvolume{13},
\bfpage{117}.
\doiurl{https://doi.org/10.1016/j.ascom.2015.10.006}.
\end{barticle}
\endbibitem

\bibitem[\protect\citeauthoryear{{Sudar}, {Vr{\v{s}}nak}, and {Dumbovi{\'c}}}{2016}]{Sudar2016}
\begin{barticle}
\bauthor{\bsnm{{Sudar}}, \binits{D.}},
\bauthor{\bsnm{{Vr{\v{s}}nak}}, \binits{B.}},
\bauthor{\bsnm{{Dumbovi{\'c}}}, \binits{M.}}:
\byear{2016},
\batitle{{Predicting coronal mass ejections transit times to Earth with neural network}}.
\bjtitle{\mnras}
\bvolume{456},
\bfpage{1542}.
\doiurl{https://doi.org/10.1093/mnras/stv2782}.
\end{barticle}
\endbibitem

\bibitem[\protect\citeauthoryear{Sundararajan, Taly, and Yan}{2017}]{DBLP:conf/icml/SundararajanTY17}
\begin{bchapter}
\bauthor{\bsnm{Sundararajan}, \binits{M.}},
\bauthor{\bsnm{Taly}, \binits{A.}},
\bauthor{\bsnm{Yan}, \binits{Q.}}:
\byear{2017},
\bctitle{Axiomatic attribution for deep networks}.
In: \bbtitle{Proceedings of the 34th International Conference on Machine Learning},
\bfpage{3319}.
\end{bchapter}
\endbibitem

\bibitem[\protect\citeauthoryear{{SunPy Community} et~al.}{2015}]{2015CS&D....8a4009S}
\begin{barticle}
\bauthor{\bsnm{{SunPy Community}}},
\bauthor{\bsnm{{Mumford}}, \binits{S.J.}},
\bauthor{\bsnm{{Christe}}, \binits{S.}},
\bauthor{\bsnm{{P{\'e}rez-Su{\'a}rez}}, \binits{D.}},
\bauthor{\bsnm{{Ireland}}, \binits{J.}},
\bauthor{\bsnm{{Shih}}, \binits{A.Y.}},
\bauthor{\bsnm{{Inglis}}, \binits{A.R.}},
\bauthor{\bsnm{{Liedtke}}, \binits{S.}},
\bauthor{\bsnm{{Hewett}}, \binits{R.J.}},
\bauthor{\bsnm{{Mayer}}, \binits{F.}},
\bauthor{\bsnm{{Hughitt}}, \binits{K.}},
\bauthor{\bsnm{{Freij}}, \binits{N.}},
\bauthor{\bsnm{{Meszaros}}, \binits{T.}},
\bauthor{\bsnm{{Bennett}}, \binits{S.M.}},
\bauthor{\bsnm{{Malocha}}, \binits{M.}},
\bauthor{\bsnm{{Evans}}, \binits{J.}},
\bauthor{\bsnm{{Agrawal}}, \binits{A.}},
\bauthor{\bsnm{{Leonard}}, \binits{A.J.}},
\bauthor{\bsnm{{Robitaille}}, \binits{T.P.}},
\bauthor{\bsnm{{Mampaey}}, \binits{B.}},
\bauthor{\bsnm{{Campos-Rozo}}, \binits{J.I.}},
\bauthor{\bsnm{{Kirk}}, \binits{M.S.}}:
\byear{2015},
\batitle{{SunPy--Python for solar physics}}.
\bjtitle{Computational Science and Discovery}
\bvolume{8},
\bfpage{014009}.
\doiurl{https://doi.org/10.1088/1749-4699/8/1/014009}.
\end{barticle}
\endbibitem

\bibitem[\protect\citeauthoryear{Tan and Le}{2019}]{Tan2019}
\begin{bchapter}
\bauthor{\bsnm{Tan}, \binits{M.}},
\bauthor{\bsnm{Le}, \binits{Q.V.}}:
\byear{2019},
\bctitle{EfficientNet: Rethinking model scaling for convolutional neural networks}.
In: \bbtitle{Proceedings of the 36th International Conference on Machine Learning}
\bseriesno{97},
\bfpage{6105}.
\end{bchapter}
\endbibitem

\bibitem[\protect\citeauthoryear{{Telloni}}{2022}]{2022FrASS...9.5880T}
\begin{barticle}
\bauthor{\bsnm{{Telloni}}, \binits{D.}}:
\byear{2022},
\batitle{{Statistical methods applied to space weather science}}.
\bjtitle{Front. Astron. Space Sci.}
\bvolume{9},
\bfpage{865880}.
\doiurl{https://doi.org/10.3389/fspas.2022.865880}.
\end{barticle}
\endbibitem

\bibitem[\protect\citeauthoryear{{Telloni} et~al.}{2023}]{2023ApJ...952..111T}
\begin{barticle}
\bauthor{\bsnm{{Telloni}}, \binits{D.}},
\bauthor{\bsnm{{Schiavo}}, \binits{M.L.}},
\bauthor{\bsnm{{Magli}}, \binits{E.}},
\bauthor{\bsnm{{Fineschi}}, \binits{S.}},
\bauthor{\bsnm{{Guastavino}}, \binits{S.}},
\bauthor{\bsnm{{Nicolini}}, \binits{G.}},
\bauthor{\bsnm{{Susino}}, \binits{R.}},
\bauthor{\bsnm{{Giordano}}, \binits{S.}},
\bauthor{\bsnm{{Amadori}}, \binits{F.}},
\bauthor{\bsnm{{Candiani}}, \binits{V.}},
\bauthor{\bsnm{{Massone}}, \binits{A.M.}},
\bauthor{\bsnm{{Piana}}, \binits{M.}}:
\byear{2023},
\batitle{{Prediction capability of geomagnetic events from solar wind data using neural networks}}.
\bjtitle{\apj}
\bvolume{952},
\bfpage{111}.
\doiurl{https://doi.org/10.3847/1538-4357/acdeea}.
\end{barticle}
\endbibitem

\bibitem[\protect\citeauthoryear{{Toriumi} and {Wang}}{2019}]{2019LRSP...16....3T}
\begin{barticle}
\bauthor{\bsnm{{Toriumi}}, \binits{S.}},
\bauthor{\bsnm{{Wang}}, \binits{H.}}:
\byear{2019},
\batitle{{Flare-productive active regions}}.
\bjtitle{Living Reviews in Solar Physics}
\bvolume{16},
\bfpage{3}.
\doiurl{https://doi.org/10.1007/s41116-019-0019-7}.
\end{barticle}
\endbibitem

\bibitem[\protect\citeauthoryear{Vourlidas, Patsourakos, and Savani}{2019}]{Vourlidas2019}
\begin{barticle}
\bauthor{\bsnm{Vourlidas}, \binits{A.}},
\bauthor{\bsnm{Patsourakos}, \binits{S.}},
\bauthor{\bsnm{Savani}, \binits{N.P.}}:
\byear{2019},
\batitle{Predicting the geoeffective properties of coronal mass ejections: current status, open issues and path forward}.
\bjtitle{Philosophical Transactions of the Royal Society A}
\bvolume{377},
\bfpage{20180096}.
\doiurl{https://doi.org/10.1098/rsta.2018.0096}.
\end{barticle}
\endbibitem

\bibitem[\protect\citeauthoryear{{Wang} et~al.}{2019}]{2019ApJ...881...15W}
\begin{barticle}
\bauthor{\bsnm{{Wang}}, \binits{Y.}},
\bauthor{\bsnm{{Liu}}, \binits{J.}},
\bauthor{\bsnm{{Jiang}}, \binits{Y.}},
\bauthor{\bsnm{{Erd{\'e}lyi}}, \binits{R.}}:
\byear{2019},
\batitle{{CME} arrival time prediction using convolutional neural network}.
\bjtitle{\apj}
\bvolume{881},
\bfpage{15}.
\doiurl{https://doi.org/10.3847/1538-4357/ab2b3e}.
\end{barticle}
\endbibitem

\bibitem[\protect\citeauthoryear{{Wanliss} and {Showalter}}{2006}]{2006JGRA..111.2202W}
\begin{barticle}
\bauthor{\bsnm{{Wanliss}}, \binits{J.A.}},
\bauthor{\bsnm{{Showalter}}, \binits{K.M.}}:
\byear{2006},
\batitle{{High-resolution global storm index: Dst versus SYM-H}}.
\bjtitle{J. Geophys. Res. (Space Physics)}
\bvolume{111},
\bfpage{A02202}.
\doiurl{https://doi.org/10.1029/2005JA011034}.
\end{barticle}
\endbibitem

\bibitem[\protect\citeauthoryear{Wilks}{2010}]{Wilks2010}
\begin{bchapter}
\bauthor{\bsnm{Wilks}, \binits{D.S.}}:
\byear{2010},
\bctitle{Forecast verification}.
In: \bbtitle{Statistical Methods in the Atmospheric Sciences},
\bedition{3}rd edn.,
\bpublisher{Elsevier}.
\bcomment{Chap. 8}.
\doiurl{https://doi.org/10.1016/B978-0-12-385022-5.00008-7}.
\end{bchapter}
\endbibitem

\bibitem[\protect\citeauthoryear{{Ye} et~al.}{2025}]{2025ApJ...978...66Y}
\begin{barticle}
\bauthor{\bsnm{{Ye}}, \binits{D.}},
\bauthor{\bsnm{{Li}}, \binits{H.}},
\bauthor{\bsnm{{Guo}}, \binits{L.}},
\bauthor{\bsnm{{Jiang}}, \binits{X.}}:
\byear{2025},
\batitle{{An improved halo coronal mass ejection geoeffectiveness prediction model using multiple coronal mass ejection features based on the DC-PCA-KNN method}}.
\bjtitle{\apj}
\bvolume{978},
\bfpage{66}.
\doiurl{https://doi.org/10.3847/1538-4357/ad98f0}.
\end{barticle}
\endbibitem

\bibitem[\protect\citeauthoryear{{Ye} et~al.}{2024}]{2024ApJ...972...52Y}
\begin{barticle}
\bauthor{\bsnm{{Ye}}, \binits{Y.}},
\bauthor{\bsnm{{Liu}}, \binits{J.}},
\bauthor{\bsnm{{Hao}}, \binits{Y.}},
\bauthor{\bsnm{{Cui}}, \binits{J.}}:
\byear{2024},
\batitle{{Evaluating the geoeffectiveness of interplanetary coronal mass ejections: Insights from a support vector machine approach with SHAP value analysis}}.
\bjtitle{\apj}
\bvolume{972},
\bfpage{52}.
\doiurl{https://doi.org/10.3847/1538-4357/ad61d7}.
\end{barticle}
\endbibitem

\bibitem[\protect\citeauthoryear{{Yu} et~al.}{2022}]{2022RAA....22f5009Y}
\begin{barticle}
\bauthor{\bsnm{{Yu}}, \binits{X.}},
\bauthor{\bsnm{{Xu}}, \binits{L.}},
\bauthor{\bsnm{{Ren}}, \binits{Z.}},
\bauthor{\bsnm{{Zhao}}, \binits{D.}},
\bauthor{\bsnm{{Sun}}, \binits{W.}}:
\byear{2022},
\batitle{Image desaturation for SDO/AIA using mixed convolution network}.
\bjtitle{Research in Astronomy and Astrophysics}
\bvolume{22},
\bfpage{065009}.
\doiurl{https://doi.org/10.1088/1674-4527/ac69b7}.
\end{barticle}
\endbibitem

\bibitem[\protect\citeauthoryear{{Zhang} et~al.}{2025}]{2025ApJ...981...37Z}
\begin{barticle}
\bauthor{\bsnm{{Zhang}}, \binits{H.}},
\bauthor{\bsnm{{Jing}}, \binits{J.}},
\bauthor{\bsnm{{Wang}}, \binits{J.T.L.}},
\bauthor{\bsnm{{Wang}}, \binits{H.}},
\bauthor{\bsnm{{Abduallah}}, \binits{Y.}},
\bauthor{\bsnm{{Xu}}, \binits{Y.}},
\bauthor{\bsnm{{Alobaid}}, \binits{K.A.}},
\bauthor{\bsnm{{Farooki}}, \binits{H.}},
\bauthor{\bsnm{{Yurchyshyn}}, \binits{V.}}:
\byear{2025},
\batitle{{Prediction of halo coronal mass ejections using SDO/HMI vector magnetic data products and a transformer model}}.
\bjtitle{\apj}
\bvolume{981},
\bfpage{37}.
\doiurl{https://doi.org/10.3847/1538-4357/adafa0}.
\end{barticle}
\endbibitem

\end{thebibliography}

\end{document}